\documentclass[12 pt, fleqn, a4paper]{article} 
\usepackage{hyperref}
\hypersetup{pdfstartview = FitH,
            pdfborder    = {0 0 0.5},
            colorlinks   = false}
 \usepackage{epsfig, subcaption}
 \usepackage{multirow}
 \usepackage{tikz}
 \usetikzlibrary{arrows.meta}
 \usepackage{clshan-math}
 \usepackage{clshan-dm-dd, clshan-dm-ddd}
 \usepackage{clshan-nu}
 \def \lsim {\:\raisebox{-0.7 ex}{$\stackrel{\textstyle <}{\sim}$}\:}

 \newcommand{\bfpN}              {{\bf p}_{\rm N}}
 \newcommand{\seta}              {\sin  (  \eta)}
 \newcommand{\sdeta}             {\sin  (2 \eta)}
 \newcommand{\sseta}             {\sin^2(  \eta)}
 \newcommand{\ssseta}            {\sin^3(  \eta)}

 \newcommand{\ceta}              {\cos  (  \eta)}
 
 \newcommand{\cceta}             {\cos^2(  \eta)}
 \newcommand{\ccceta}            {\cos^3(  \eta)}

 \newcommand{\spsi}              {\sin  (  \psi)}
 
 \newcommand{\ssspsi}            {\sin^3(  \psi)}
 \newcommand{\cpsi}              {\cos  (  \psi)}

 \newcommand{\stheta}            {\sin  (  \theta_{\rm N_R, \nu_{in}})}
 \newcommand{\sdtheta}           {\sin  (2 \theta_{\rm N_R, \nu_{in}})}
 \newcommand{\sstheta}           {\sin^2(  \theta_{\rm N_R, \nu_{in}})}

 \newcommand{\cctheta}           {\cos^2(  \theta_{\rm N_R, \nu_{in}})}
 \newcommand{\ccctheta}          {\cos^3(  \theta_{\rm N_R, \nu_{in}})}

 \newcommand{\FSISDQ}            {F_{\rm (SI, SD)}^2 (Q)}
 \def \PeriodAa        {  0    -- 365}
 \def \PlotNumberAa    {00000}
 \def \PeriodCa        { 19.49 --  79.49}
 \def \PeriodCb        {110.74 -- 170.74}
 \def \PeriodCc        {201.99 -- 261.99}
 \def \PeriodCd        {293.24 -- 353.24}
 \def \PlotNumberCa    {04949}
 \def \PlotNumberCb    {14074}
 \def \PlotNumberCc    {23199}
 \def \PlotNumberCd    {32324}
 \newcommand{\rmB}               {\rmXA{B}   {8}}
 \newcommand{\rmF}     [1] [ 19] {\rmXA{F}  {#1}}
 \newcommand{\rmAr}              {\rmXA{Ar} {40}}
 \newcommand{\rmGe}    [1] [ 73] {\rmXA{Ge} {#1}}
 \newcommand{\rmXe}    [1] [129] {\rmXA{Xe}{129}}
 \newcommand{\rmW}               {\rmXA{W} {183}}
 \newcommand{\rmS}               {\rmXA{S}  {32}}
 \newcommand{\rmCl}    [1] [ 35] {\rmXA{Cl} {#1}}
 \newcommand{\rmBr}    [1] [ 79] {\rmXA{Br} {#1}}
 \newcommand{\rmI}               {\rmXA{I} {127}}
\newcommand{\InsertSKPPlotS} [3] [8.25] {
\begin{figure} [t!]
\begin{center}
 \includegraphics [width = #1 cm] {skp-#2}
\end{center}
\caption{
 #3
}
\label{fig:#2}
\end{figure}
}
\newcommand{\InsertSKPPlotD} [6] {
\begin{figure} [t!]
\begin{center}
 \begin{subfigure} [c] {8.25 cm}
  \includegraphics [width = 8.25 cm] {skp-#1}
 \caption{#3}
 \end{subfigure}
 \hspace{0.1 cm}
 \begin{subfigure} [c] {8.25 cm}
  \includegraphics [width = 8.25 cm] {skp-#2}
 \caption{#4}
 \end{subfigure}
\end{center}
\caption{
 #6
}
\label{fig:#5}
\end{figure}
}
\newcommand{\InsertPlotNang} [5] {%
 \begin{minipage} {#1 cm}
  {\begin{center}
    \includegraphics [width = #1 cm]
                     {#2_ang-#3-\ShortFrame-\EventNumber-#4} \\ \vspace{0.05 cm}
    {\small #5}
   \end{center}}
 \end{minipage}
}
\newcommand{\InsertSubfigureNangAnnual} [3] {
 \InsertPlotNang
  {8}
  {#1} {#2} {\PlotNumberAa#3} {\PeriodAa\ day}%
 \\ \vspace{0.2 cm}
 \InsertPlotNang
  {4}
  {#1} {#2} {\PlotNumberCa#3} {\PeriodCa\ day}%
 \InsertPlotNang
  {4}
  {#1} {#2} {\PlotNumberCb#3} {\PeriodCb\ day}%
 \InsertPlotNang
  {4}
  {#1} {#2} {\PlotNumberCc#3} {\PeriodCc\ day}%
 \InsertPlotNang
  {4}
  {#1} {#2} {\PlotNumberCd#3} {\PeriodCd\ day}%
}
\newcommand{\InsertPlotNR} [4] {
 \begin{subfigure} [c] {#1 cm}%
  \includegraphics [width = #1 cm]
                   {#2-\Target-#3-\EnergyWindow-\EventNumber-#4}%
 \end{subfigure}
}
\newcommand{\InsertPlotNRtheta} [2] {
 \InsertPlotNR
  {8.4} {#1} {#2} {\PlotNumber}%
}
\newcommand{\InsertFigureNRtheta} [3] {
\begin{figure} [t!]
\begin{center}
 \InsertPlotNRtheta {#1NR_ang}    {#2-\ShortFrame}%
 \InsertPlotNRtheta {#1NR_theta}  {#2}%
 \\ \vspace{0.6 cm}
 \InsertPlotNRtheta {#1Q_ang}     {#2-\ShortFrame}%
 \InsertPlotNRtheta {#1Q_theta}   {#2}%
 \\ \vspace{0.6 cm}
 \InsertPlotNRtheta {#1QoN_ang}   {#2-\ShortFrame}%
 \InsertPlotNRtheta {#1QoN_theta} {#2}%
\end{center}
\caption{
  #3
}
\label{fig:#1NR_theta-\Target-#2-\EnergyWindow-\EventNumber-\PlotNumber}
\end{figure}
}
\newcommand{\InsertSubfigureNRangAnnual} [3] {
 \InsertPlotNR
  {5}
  {#1_ang} {#2-\ShortFrame} {\PlotNumberAa#3}%
 \\ \vspace{0.1 cm}
 \InsertPlotNR
  {4}
  {#1_ang} {#2-\ShortFrame} {\PlotNumberCa#3}%
 \InsertPlotNR
  {4}
  {#1_ang} {#2-\ShortFrame} {\PlotNumberCb#3}%
 \InsertPlotNR
  {4}
  {#1_ang} {#2-\ShortFrame} {\PlotNumberCc#3}%
 \InsertPlotNR
  {4}
  {#1_ang} {#2-\ShortFrame} {\PlotNumberCd#3}%
}
\newcommand{\InsertFigureNRangAnnual} [6] {
\begin{figure} [t!]
\begin{center}
 \InsertSubfigureNRangAnnual {#1} {#4} {#5}%
 \\ \vspace{0.5 cm}
 \InsertSubfigureNRangAnnual {#2} {#4} {#5}%
 \\ \vspace{0.5 cm}
 \InsertSubfigureNRangAnnual {#3} {#4} {#5}%
\end{center}
\caption{
 #6
}
\label{fig:#1_ang-\Target-#4-\ShortFrame-\EnergyWindow-\EventNumber#5}
\end{figure}
}
\begin{document}
\thispagestyle{empty}
\begin{flushright}
 June 2026
\end{flushright}
\begin{center}
{\Large\bf
 Simulations of 3-Dimensional Recoil Response to      \\
 Coherent Elastic Neutrino--Nucleus Scattering Events \\ \vspace{0.2 cm}
 in Directional Direct Dark Matter Detectors}         \\
\vspace*{0.7 cm}
 {\sc Kwang-Chang Lai}$^{\ddagger}$
 and
 {\sc Chung-Lin Shan}$^{\S}$                          \\~\\
\vspace{0.5 cm}
 ${}^{\ddagger}$%
 {\small\it
  Center for General Education,
  Chang Gung University                               \\ \vspace{0.05 cm}
  Kwei-Shan, Taoyuan, 333323, Taiwan, R.O.C.}         \\ \vspace{0.5  cm}
 {\it E-mail:}
 {\tt kcl@mail.cgu.edu.tw}                            \\ \vspace{1.5  cm}
 ${}^{\S}$%
 {\small\it
  Preparatory Office of
  the Supporting Center for
  Taiwan Independent Researchers                       \\ \vspace{0.05 cm}
  P.O.BOX 21 National Yang Ming Chiao Tung University,
  Hsinchu City 300093, Taiwan, R.O.C.}                 \\ \vspace{0.5  cm}
 {\it E-mail:}
 {\tt clshan@tir.tw}
\end{center}
\vspace{2 cm}
\begin{abstract}

 Following our earlier work on studying
 3-dimensional nuclear recoil response to
 Galactic Weakly Interacting Massive Particles (WIMPs)
 in directional direct Dark Matter detectors,
 in this paper,
 we simulate
 3-D coherent elastic neutrino--nucleus scattering (\CEnuNS) events
 induced by Solar $\rmB$ neutrinos.
 Our numerical results show that,
 in contrast to
 the approximately fixed patterns of
 the WIMP--induced signals,
 the characteristic ring--like
 angular distributions of
 the nuclear recoil flux/energy of
 \CEnuNS\ events
 show clearly annual variations
 along the trajectories of
 the moving direction of
 incident Solar neutrinos
 in different celestial coordinate systems
 without experimentally distinguishable target dependence.

\end{abstract}
\clearpage
 \tableofcontents
 \addtocontents{toc}{}
 \clearpage
\section{Introduction}

 A half--century ago,
 D.~Z.~Freedman proposed the possibility of
 elastic scattering of low--energy neutrinos
 off atomic nuclei
 via the neutral electroweak current
 \cite{Freedman74, Drukier84}.
 The estimated recoil energy
 would be maximal a few tens of keV
 and down to the sub--keV range.
 Since
 the transferred momenta
 from incident neutrinos
 are so low that
 the associated wavelengths
 are comparable to or larger than
 the size of the nucleus,
 the scattering amplitudes from individual nucleons add coherently.
 This coherent scattering
 enlarges significantly the cross section,
 approximately proportional to
 the square of the neutron number of
 the target nucleus.

\begin{figure} [t!]
\begin{center}
 \includegraphics [width = 10.5 cm] {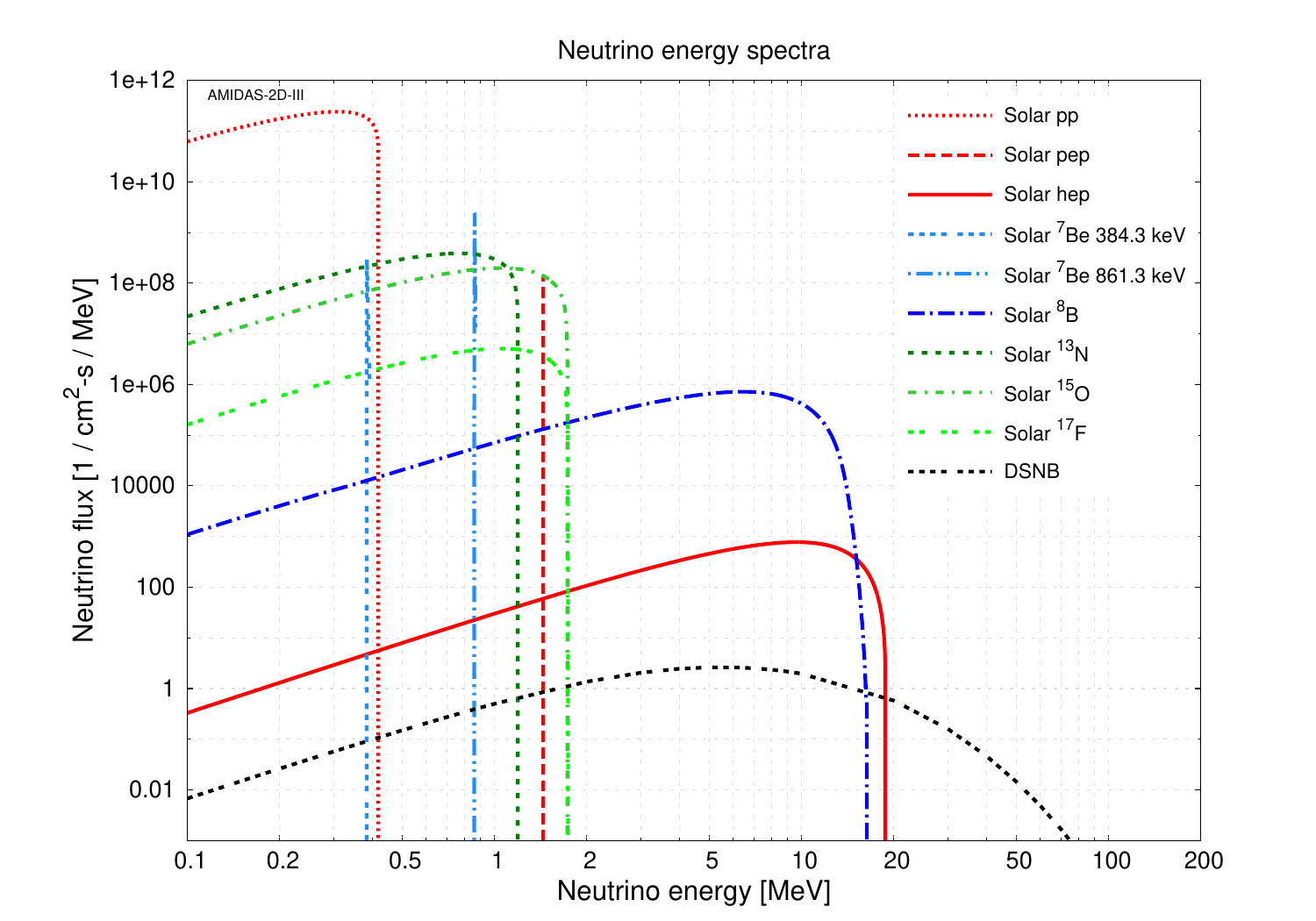}
\end{center}
\caption{
 Theoretically estimated neutrino flux
 for each type of Solar neutrinos.
 See the text for further details.
}
\label{fig:dPhidEnu}
\end{figure}

\begin{figure} [b!]
\begin{center}
 \begin{subfigure} [c] {8.25 cm}
  \includegraphics [width = 8.25 cm] {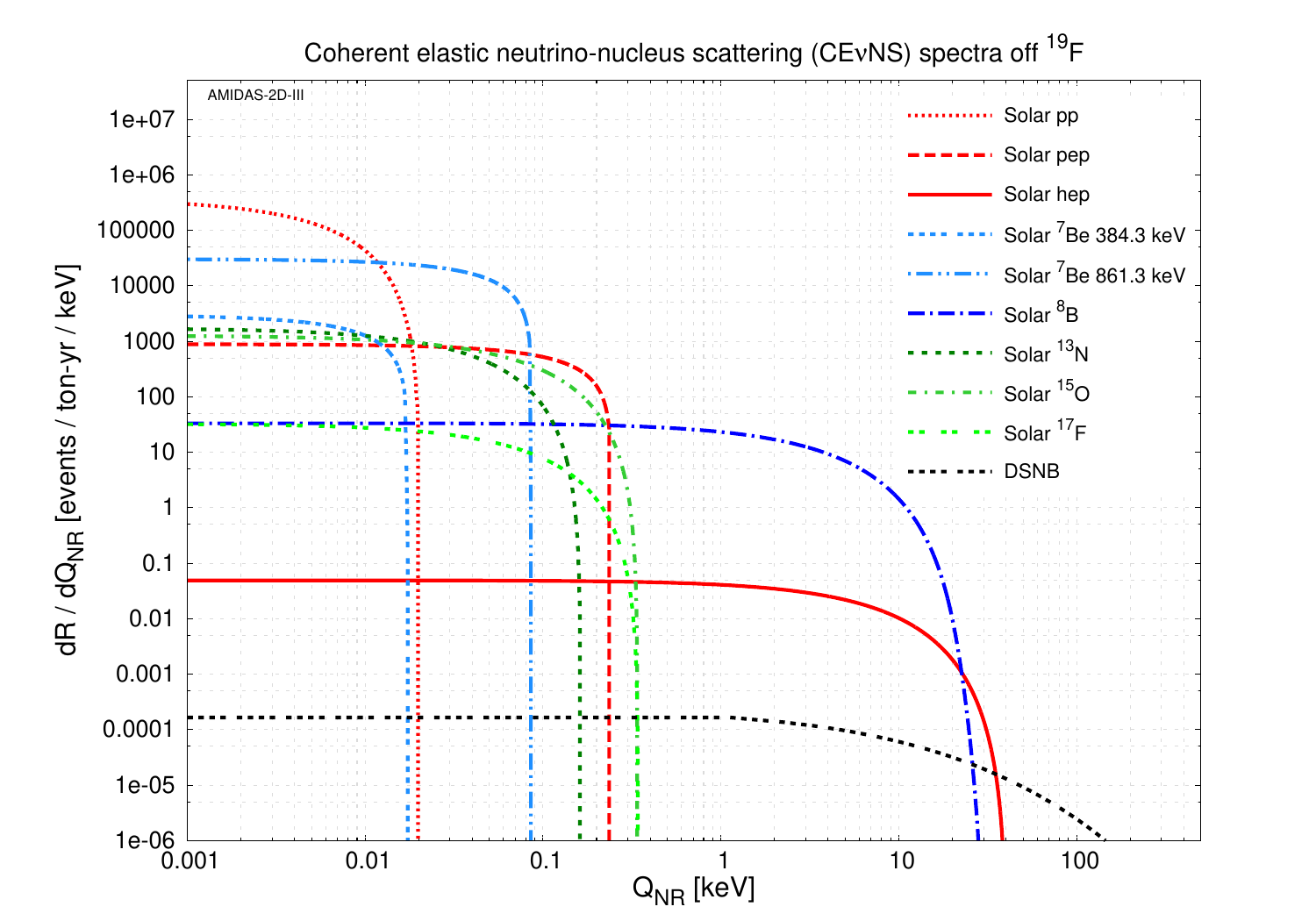}
 \caption{Off $\rmF$}
 \end{subfigure}
 \hspace{0.1 cm}
 \begin{subfigure} [c] {8.25 cm}
  \includegraphics [width = 8.25 cm] {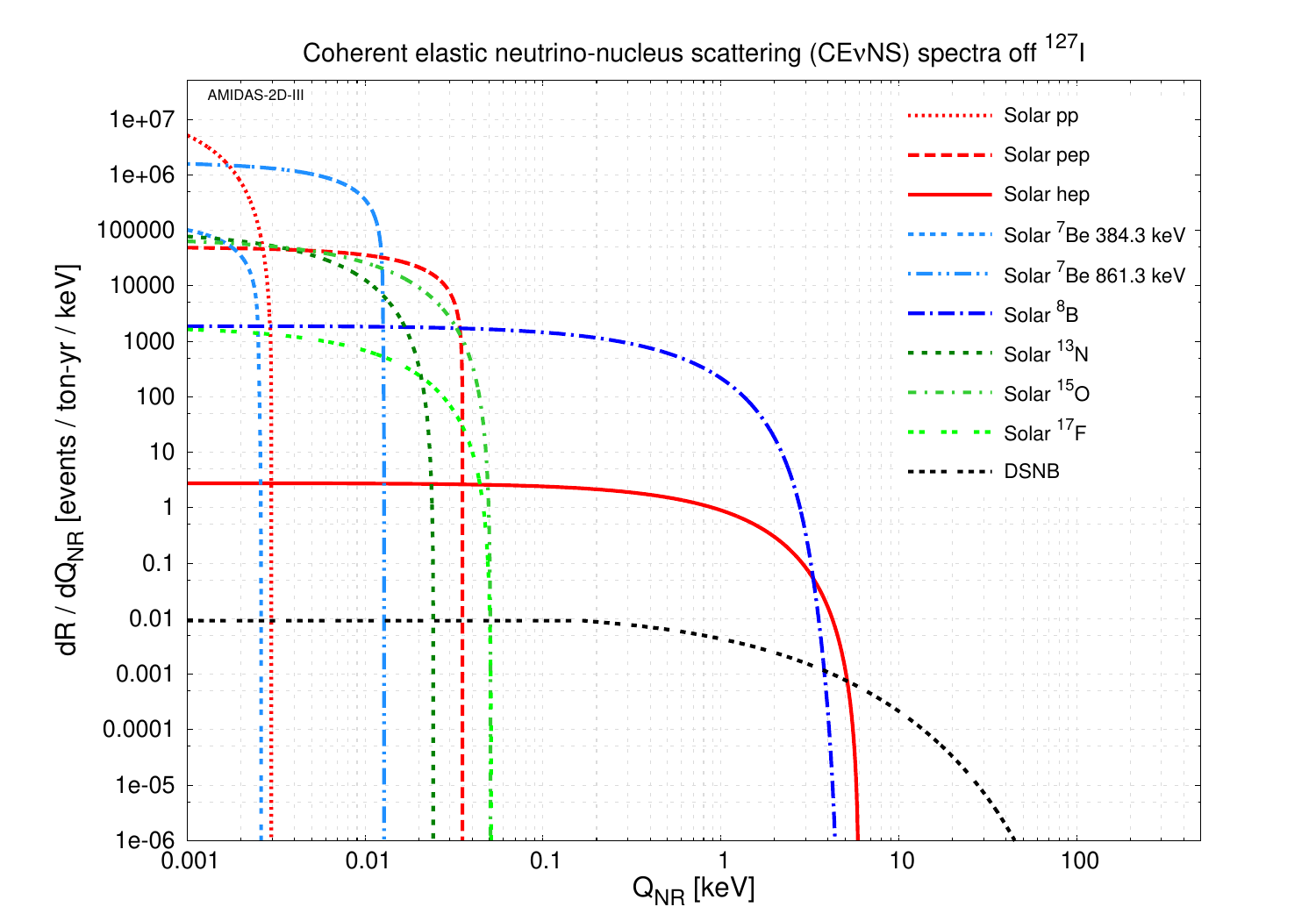}
 \caption{Off $\rmI$}
 \end{subfigure}
\end{center}
\caption{
 Theoretically estimated
 \CEnuNS\ event rate
 as a function of the recoil energy
 for each type of Solar neutrinos
 shown in Fig.~\ref{fig:dPhidEnu}
 off (a) $\rmF$ and (b) $\rmI$ nuclei,
 respectively.
}
\label{fig:dRnuNSMdQ}
\end{figure}

 Among various neutrino sources,
 Solar neutrinos play a dominant role.
 In particular,
 neutrinos produced in the $\beta$--decay of $\rmB$ nuclei
 in the Solar core
 possess the highest energies among Solar neutrino components.
 In Fig.~\ref{fig:dPhidEnu},
 we show
 the theoretically estimated neutrino flux
 as a function of the neutrino energy
 for each type of Solar neutrinos.
 The normalized energy probability distributions of
 different types of Solar neutrinos
 are adopted from Ref.~\cite{Solarnuspectrum}
 and multiplied by the fluxes
 estimated by the Barcelona-2016 calculation
 with the Grevesse--Sauval--1998 Standard Solar Model data set
 (B16-GS98)
 listed in Refs.~%
 \cite{Vinyoles:1611.09867,
       Nakahata:2202.12421, SChen:2501.09971}%
\footnote{
 The accompanied spectrum of
 the diffuse supernova neutrino background
 is estimated numerically
 based on the analytical expression and parameter values
 given in Ref.~\cite{DeRomeri:2309.04117}.
 The temperatures of
 different flavors
 are set as:
  6.6 MeV for electron neutrinos,
  7.0 MeV     electron anti--neutrinos,
 and 10.0 MeV muon and tau neutrinos,
 respectively.
}.
 One can see clearly that,
 except for
 the $\rmB$ and hep neutrinos,
 the incident energy of most types of Solar neutrinos
 are less than 2 MeV.
 Hence,
 as
 the theoretically estimated
 coherent elastic neutrino--nucleus scattering (\CEnuNS) spectrum
 for each type of Solar neutrinos
 off (a) $\rmF$ and (b) $\rmI$ nuclei
 shown correspondingly in Figs.~\ref{fig:dRnuNSMdQ}
 indicate,
 the maximal transferrable recoil energy of
 target nuclei as light as $\rmF$
 is less than 0.35 keV.
 For heavy target nuclei like $\rmI$,
 the maximal recoil energy
 can even be as low as only 0.05 keV.

 Fig.~\ref{fig:dPhidEnu} shows also that,
 the ratio between
 the maximal fluxes of
 the two most energetic Solar ($\rmB$ and hep) neutrinos
 is $\sim 10^3$.
 This means that,
 as long as
 we consider
 only a few tens to hundreds of \CEnuNS\ events,
 it should be safe to only take into account
 the Solar $\rmB$ neutrinos.
 Due to the same reason,
 we ignore
 the diffuse supernova neutrino background (DSNB) events,
 whose maximal flux is
 two orders of magnitude smaller than
 that of the Solar hep neutrinos,
 although
 these neutrinos
 are expectedly much more energetic than 20 MeV
 (and even up to \mbox{$\sim$ 1 TeV}).


 On the other hand,
 Weakly Interacting Massive Particles (WIMPs)
 arising in several extensions of
 the Standard Model of particle physics
 are one of the most favorite candidates
 for cosmological Dark Matter (DM).
 In the last four decades,
 a large number of experiments has been built
 and is being planned
 to search for different WIMP candidates
 by
 direct detection of
 coherent (elastic) WIMP--nucleus scattering
 observed
 in ultra--low background detectors
 at deep underground laboratories.
 These would be
 the most reliable experimental strategy
 for identifying Galactic DM particles
 and determining their properties
  \cite{SUSYDM96, Schumann19, Baudis20, Cooley21, Cirelli24, Baudis25}.

 Over the past two decades,
 great progresses in detector mass and purity,
 background shielding and rejection
 as well as
 energy threshold and signal readout
 have
 let us be able to explore deeper and deeper area
 in the WIMP parameter space,
 corresponding to a much lowered scattering event rate
 \cite{ZBo24, Aalbers24d, Aprile25a}.
 As a result,
 the above mentioned
 neutrino--nucleus scattering events
 would on one hand be detectable
 by using future generations of
 direct DM/WIMP detectors
 \cite{Strigari09a,
       ZBo:2407.10892, Aprile:2408.02877, Akerib:2512.08065, Aprile:2604.06002}.
 However,
 this also means that
 Galactic WIMPs and Solar neutrinos
 become mutual unignorable background%
\footnote{
 In direct Dark Matter detection physics,
 the predicted limits/areas
 with light ($\lsim {\cal O}(10~{\rm GeV})$) WIMP masses
 in the WIMP mass versus cross section planes
 are called the neutrino floor or the neutrino fog
 \cite{OHare16, Papoulias18, OHare20, OHare21,
       AristizabalSierra21, Bloch24}.
}
 \cite{Guetlein10,
       Billard13b, Gaspert21}.
 As discussed in the literature
 \cite{Billard13b,
       OHare15b}
 and shown later,
 once the WIMP mass is below ${\cal O}(10~{\rm GeV})$,
 its theoretically predicted scattering spectrum
 could almost perfectly mimic the \CEnuNS\ spectrum.

 Nevertheless,
 while
 most direct Dark Matter detection experiments
 measure only recoil energies,
 the directional direct detectors
 proposed more than one decade ago
 are aimed to provide
 additional 3-dimensional information
 (recoil tracks and/or head--tail senses)
 of (WIMP--)nucleus scattering events
 \cite{Ahlen09, Mayet16, Vahsen21}.
 Several experimental collaborations
 investigating different detector techniques and materials
 \cite{Agafonova17, Ikeda20, Marshall20}
 have also achieved recently great progress
 \cite{Mayet16, Battat16b,
       Vahsen20,
       Miuchi23}.
 Due to
 the orbital motion of
 the Solar system
 through the Galactic halo,
 the incident flux of
 halo DM/WIMPs
 is considered to be peaked around
 the direction opposite to
 the Solar Galactic movement,
 namely,
 the direction approximately
 from the Cygnus constellation to the Solar center,
 while
 Solar neutrinos come directly
 from the Solar center to the Earth's center.
 Hence,
 directional--sensitive direct detection
 could be a promising experimental strategy
 for not only detecting but also discriminating
 WIMP and neutrino induced signals
 from each other
 \cite{Ruppin14,
       OHare15b, OHare16, OHare20,
       Abdullah20,
       YZhuang23, YZhuang24,
       Lisotti24, AristizabalSierra24, Shekar25,
       OHare26},

 In our earlier work
 \cite{DMDDD-NR},
 we studied
 the 3-dimensional nuclear recoil flux and energy distributions of
 WIMP--induced scattering signals
 in directional direct DM detectors
 in a wide WIMP mass range.
 In this paper,
 we focus on
 the detector response to
 the neutrino--nucleus scattering events
 induced by Solar $\rmB$ neutrinos
 in different observation periods%
\footnote{
 Detailed investigations on
 the 3-D nuclear recoil response to
 mixed Galactic WIMP and Solar $\rmB$ neutrino scattering signals
 in directional direct DM detectors
 with different target nuclei
 is currently in finalization
 and will be announced soon.
}.
 Instead of
 the expression for
 the differential event rate of \CEnuNS\
 in many previous works
 explored by other authors
 \cite{Strigari09a,
       Billard13b, Ruppin14,
       OHare15b,
       OHare16, OHare20,
       Abdullah20,
       Lisotti24, AristizabalSierra24,
       Shekar25,
       OHare26},
 we simulate
 the elastic neutrino--nucleus scattering process
 in a microscopic scattering--by--scattering approach
 based on classical elastic two--body collision
 (described in detail in Sec.~\ref{subsec:3D-CEnuNS}).
 This would
 not only
 be easily extended to superpose scattering events
 induced by particles
 incoming from different directions in different time,
 but also
 reproduce the physical mechanism
 more realistically.

 The remainder of this paper is organized as follows.
 In Sec.~2,
 we give
 a theoretical analysis of recoil kinematics of
 3-dimensional coherent elastic neutrino--nucleus scattering
 and derive
 the expressions for
 the angular distributions of
 the nuclear recoil flux and energy of
 3-D \CEnuNS\ events.
 Then,
 in Sec.~3,
 we describe
 the overall workflow of
 our double--Monte Carlo (MC) scattering--by--scattering simulation process of
 3-D \CEnuNS\ events,
 the first MC generation of
 the 3-D information
 (the energy,
  the incoming/scattering time
  and thus the moving direction)
 of incident Solar $\rmB$ neutrinos
 as well as
 the second MC generation of
 the 3-D information
 (the recoil energy
  and the moving direction)
 of the scattered target nucleus
 and
 the validation of
 each generated
 3-D \CEnuNS\ event.
 Numerical simulation results
 will be presented in Sec.~4.
 We summarize our observations
 in Sec.~5.
 Some technical details for our simulation and analysis
 are collected in Appendix.

\section{Formalism}
\label{sec:formalism}

 In this section,
 we discuss
 the recoil kinematics of
 3-dimensional coherent elastic neutrino--nucleus scattering
 and derive
 the expression for
 the angular distributions of
 the nuclear recoil flux and energy of
 3-D \CEnuNS\ events.
 The latter will be used
 as the most important validation criterion
 in our Monte Carlo scattering--by--scattering simulation.

\subsection{Differential scattering cross section}
\label{subsec:CEnuNS:SM:dsigma_dQ}
\begin{table} [t!]
\small
\begin{center}
\renewcommand{\arraystretch}{1.85}
\begin{tabular}{|| c   c   c   c   c   c ||}
\hline
\hline
 \makebox[2  cm][c]{Isotope}        &
 \makebox[1.5cm][c]{$Z$}            & \makebox[1.5cm][c]{$J$}     &
 \makebox[2.5cm][c]{$\Srmp$}        & \makebox[2.5cm][c]{$\Srmn$} &
 \makebox[4.5cm][c]{Natural abundance (\%)} \\
\hline
\hline
 $\rmXA{F}{19}$   &  9 & 1/2 &                    0.441   &  \hspace{-1.8ex}$-$0.109   &
 100     \\
\hline
 $\rmXA{Cl}{35}$  & 17 & 3/2 &  \hspace{-1.8ex}$-$0.059   &  \hspace{-1.8ex}$-$0.011   &
  75.78  \\
\hline
 $\rmXA{Cl}{37}$  & 17 & 3/2 &  \hspace{-1.8ex}$-$0.058   &                    0.050   &
  24.22  \\
\hline
 $\rmXA{I}{127}$  & 53 & 5/2 &                    0.309   &                    0.075   &
 100     \\
\hline
\hline
\end{tabular}
\end{center}
\caption{
 List of
 the commonly adopted group spin values of
 the nuclei
 used in directional direct DM detection experiments
 \cite{Engel95, Ressell97,
       Bednyakov04a,
       Giuliani05,
       WebElements}.
}
\label{tab:Sp/n}
\end{table}

 The general expression for
 the differential coherent elastic neutrino--nucleus scattering
 cross section
 in the Standard Model (SM)
 has been given by
 \cite{Strigari09a,
       Guetlein10,
       Ruppin14,
       Papoulias18, Abdullah20,
       Barbeau21}%
\footnote{
 Note here that
 $\FSISDQ$ indicates
 the nuclear form factors
 corresponding to
 the spin--independent vector
 and the spin--dependent axial--vector interactions,
 normalized to $F_{\rm (SI, SD)} (0) = 1$,
 respectively.
}
\beqn
     \Dd{\sigmanuNSM}{Q}
 \=  \afrac{G_F^2}{2 \pi} \mN
     \bbrac{  \aBig {\grmv   + \grma  }^2
            + \aBig {\grmv   - \grma  }^2 \abrac{1 - \frac{Q}{\Enu}}^2
            - \abrac{\grmv^2 - \grma^2}   \afrac{\mN Q}{\Enu^2}         }
     \FSISDQ
     \non\\
 \eqnapprox
     \afrac{G_F^2}{2 \pi} \mN
     \bbrac{  \grmv^2 \abrac{2 - \frac{\mN Q}{\Enu^2}} \FSIQ
            + \grma^2 \abrac{2 + \frac{\mN Q}{\Enu^2}} \FSDQ  }
     \non\\
 \eqnapprox
     \afrac{G_F^2}{\pi} \grmv^2 \mN
     \abrac{1 - \frac{\mN Q}{2 \Enu^2}} \FSIQ
\~.
\label{eqn:dsigma_nuN_dQQ}
\eeqn
 Here
 we have
 the nuclear recoil energy
 $Q    \~ \lsim \~ {\cal O}(100     ~{\rm keV})$,
 the energy of incident neutrino
 $\Enu \~ \lsim \~ {\cal O}(100     ~{\rm MeV})$,
 the mass of the target nucleus
 $\mN  \~ \lsim \~ {\cal O}(20 - 200~{\rm GeV})$,
 and
 the Fermi constant
 \cite{RPP24Const}
\beq
     \frac{G_F}{(\hbar c)^3}
  =  1.166~378~5(6) \times 10^{-5}~{\rm GeV}^{-2}
\~.
\label{eqn:G_F}
\eeq
 Moreover,
 the effective spin--independent (SI) vector
 and spin--dependent (SD) axial--vector couplings of a nucleus
 to the Z$^0$ boson
 $\grmv$ and $\grma$
 are given by
 \cite{Papoulias18}
\cheqna
\beqn
     \grmv
 \=  N \grmvn + Z \grmvp
     \non\\
 \eqnsim
   - \frac{1}{2} \cbigg{N -   \bbrac{1 - 4 \ssthetaW} Z}
     \non\\
 \=- \frac{1}{2} \~ \aBig{N - 0.1078 \~ Z}
\~,
\label{eqn:grmv}
\eeqn
 and
\cheqnb
\beqn
     \grma
 \=  \abrac{N_{+} - N_{-}} \grman
   + \abrac{Z_{+} - Z_{-}} \grmap
     \non\\
 \eqnsim
     \abrac{N_{+} - N_{-}}
     \abrac{-\frac{1}{2}}
   + \abrac{Z_{+} - Z_{-}}
     \abrac{\frac{1}{2}}
     \non\\
 \eqnBinary{=}
   - \Srmn + \Srmp
\~,
\label{eqn:grma}
\eeqn
\cheqn
 respectively.
 Here
 $A$ is the nuclear mass number,
 $Z$ is the atomic/proton number
 and
 thus
 $N = A - Z$ is the neutron number;
 $N_{\pm}$ and $Z_{\pm}$
 are the numbers of the neutrons and the protons
 with the spin of $\pm \frac{1}{2}$,
 respectively,
 and
 $\expv{S_{\rm (n, p)}}$ are thus
 the expectation values of the neutron and proton group spins,
 respectively.
 In Table \ref{tab:Sp/n},
 we list the commonly adopted group spin values of
 the nuclei
 used in directional direct DM detection experiments.
 In addition,
 the vector and axial--vector couplings on neutrons and on protons are
 \cite{Sahu20, Barbeau21}
\cheqna
\beq
       \grmvn
 \sim
   - \frac{1}{2}
\~,
     ~~~~ ~~~~ ~~~~ ~~~~ 
       \grmvp
 \sim
     \frac{1}{2} - 2 \ssthetaW
\~,
\label{eqn:grmvn-grmvp}
\eeq
 and
\cheqnb
\beq
       \grman
 \sim
   - \frac{1}{2}
\~,
     ~~~~ ~~~~ ~~~~ ~~~~ 
       \grmap
 \sim
     \frac{1}{2}
\~,
\label{eqn:grman-grmap}
\eeq
\cheqn
 respectively,
 and
 $\thetaW$
 is the weak mixing angle
 with
 $\ssthetaW = 0.23122$
 \cite{RPP24Const}.

 Since
 the absolute values of
 the differences between the neutron and the proton group spins
 are less than 1,
 the SD interaction
 in Eq.~(\ref{eqn:dsigma_nuN_dQQ})
 is typically smaller by a factor of $4 / N^2$.
 The differential \CEnuNS\ cross section
 $d\sigmanuNSM / dQ$
 would therefore be dominated
 by the SI interaction.

\subsection{Nuclear recoil energy}
\label{subsec:CEnuNS:SM:QQ}

\begin{figure} [t!]
\begin{center}
\begin{tikzpicture}
      [vector/.style = {-Stealth,
                        line width = 2 pt},
       dashed/.style = {dash pattern = on 0.225 cm off 0.125 cm,
                        line width = 1.5 pt},
       angle/.style  = {Stealth-Stealth, line width = 0.75 pt}]
 \draw [very thick, color = white]
       (0, 0) rectangle (13   ,  6.5 );
%
%
 \draw [fill]      (  1.5  , 3    ) circle [radius = 0.1 ]
                  +(  0    ,-0.1  ) [below] node {\large $\nu$};
 \draw [vector]    (  1.75 , 3    ) -- (  6.75 , 3    );
 \node [below]  at (  4.25 , 2.9  ) {\large $\bfpnu$};
%
%
 \draw [fill]      (  7    , 3    ) circle [radius = 0.15]
                  +(- 0.15 , 0.1  ) [above] node {\large $\mN$};
%
%
 \draw [vector]    (  7    , 3    ) +( 0.263, 0.425) -- +( 2.167, 3.5  );
 \node [below]  at (  7.75 , 5.5  ) {\large $\bfpnu'$};
 \draw [angle]     (  7.75 , 3    ) arc [radius = 0.75 , start angle =   0, end angle =  58.237];
 \node          at (  7.9  , 3.55 ) {\large $\zeta$};
%
%
 \draw [vector]    (  7    , 3    )                  -- +( 1.25 ,-1.5  );
 \node [below]  at (  8.65 , 1.7  ) {\large $\bfpN'$};
 \draw [angle]     (  7.75 , 3    ) arc [radius = 0.75 , start angle =   0, end angle = -50.194];
 \node          at (  7.9  , 2.45 ) {\large $\eta$};
%
%
 \draw [dashed]    (  7.25 , 3    ) -- +( 3.025, 0    );
 \node [right]  at ( 10.3  , 2.9  ) {\large $\znu$};
%
 \draw [dashed]    (  7    , 2.75 ) -- +( 0    ,-2.325);
 \node [below]  at (  7.5  , 0.4  ) {\normalsize $\xnu$--$\ynu$ plane};
 \draw [angle]     (  7    , 1.95 ) arc [radius = 1.05, start angle = -90, end angle = -50.194];
 \node          at (  7.65 , 1.55 ) {\normalsize $\thetaNRnu$};
%
%
%
 \draw [dashed]    (  7    , 3    ) +( 1.25 ,-1.5  ) -- +( 2.167, 3.5  );
 \draw [angle]     (  9.175, 3    ) arc [radius = 0.65, start angle =   0, end angle =  82.683];
 \node          at (  9.2  , 3.6  ) {\large $\psi$};
\end{tikzpicture}
\end{center}
\caption{
 The relation between
 the scattering angle of
 the outgoing neutrino
 $\zeta$,
 the recoil angle of
 the scattered target nucleus
 $\eta$,
 and
 the elevation of
 the recoil direction of
 the target nucleus
 (the equivalent recoil angle)
 $\thetaNRnu$
 in the incoming--neutrino coordinate system,
 in which
 the $\znu$--axis is defined as
 the direction of
 the incoming momentum of
 the incident neutrino of interest
 $\bfpnu$.
}
\label{fig:nu-N}
\end{figure}
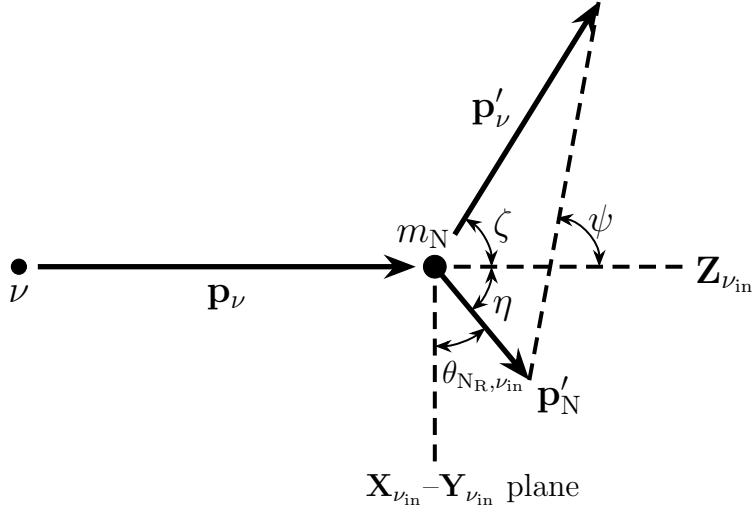

 In Fig.~\ref{fig:nu-N},
 we sketch
 the relation between
 the scattering angle of
 the outgoing neutrino
 $\zeta$,
 the recoil angle of
 the scattered target nucleus
 $\eta$,
 and
 the elevation of
 the recoil direction of
 the target nucleus
 (the equivalent recoil angle)
 $\thetaNRnu$
 in the incoming--neutrino ($\nuin$) coordinate system%
\footnote{
 This is conventionally called
 the laboratory (Lab) reference frame,
 which has however been used
 for defining the coordinate system
 fixed to our laboratory/detector of interest
 in our simulation package.
 See Sec.~\ref{subsubsec:XYZ_nu}
 for detailed definition of
 the incoming--neutrino coordinate system.
},
 in which
 the $\znu$--axis is defined as
 the direction of
 the incoming momentum of
 the incident neutrino of interest
 $\bfpnu$.
 Fig.~\ref{fig:nu-N} shows that
 $\thetaNRnu$
 is namely the complementary angle of
 the recoil angle $\eta$:
\beq
      \thetaNRnu
  =   \frac{\pi}{2} - \eta
  =   \frac{\psi}{2}
\~.
\label{eqn:thetaNRnu_eta}
\eeq
 Here
 $\psi$
 is the scattering angle of
 the outgoing neutrino
 in the center--of--momentum (CM) reference frame.

\subsubsection[In the $\nuin$ coordinate system]
              {\boldmath
               In the $\nuin$ coordinate system}
\label{subsubsec:CEnuNS:SM:QQ:eta}
\label{subsubsec:CEnuNS:SM:QQ:theta}

 The nuclear recoil energy
 induced by \CEnuNS\
 can be given exactly by
 the recoil angle
 in the incoming--neutrino coordinate system
 as
 \cite{Abdullah20}
\beqn
     Q (\Enu, \eta)
 \=  \frac{2 \mN \Enu^2 \cceta}{(\mN + \Enu)^2 - \Enu^2 \cceta}
     \non\\
 \eqnapprox
     2 \mN \afrac{\Enu}{\mN + \Enu}^2 \cceta
     \non\\
 \eqnapprox
     \afrac{2 \Enu^2}{\mN} \cceta
\~.
\label{eqn:QQ_eta}
\eeqn
 This gives,
 for a given neutrino incident energy $\Enu$,
 the maximal transferable recoil energy
 \cite{Drukier84,
       Strigari09a,
       Guetlein10,
       Harnik12,
       Billard13b}:
\beqn
     \Qmax (\Enu)
  =  Q (\Enu, \eta = 0)
 \=  \frac{2 \Enu^2}{\mN + 2 \Enu}
 \approx
     \frac{2 \Enu^2}{\mN}
\~,
\label{eqn:Qmax_Enu}
\eeqn
 as well as
\cheqna
\beqn
     \Dd{Q(\Enu, \eta)}{\eta}
 \=- \frac{(2 \mN + Q)^2}{2 \mN}
     \afrac{\Enu}{\mN + \Enu}^2 \sdeta
     \non\\
 \eqnapprox
   - 2 \mN
     \afrac{\Enu}{\mN + \Enu}^2 \sdeta
     \non\\
 \eqnapprox
   - \afrac{2 \Enu^2}{\mN} \sdeta
\~,
\label{eqn:dQQ_deta}
\eeqn
 and
\cheqnb
\beqn
     \Dd{Q (\Enu, \eta)}{\Omega_{\eta}}
 \propto
      \frac{1}{\seta} \cdot \Dd{Q (\Enu, \eta)}{\eta}
 \=- \frac{(2 \mN + Q)^2}{\mN}
     \afrac{\Enu}{\mN + \Enu}^2 \ceta
     \non\\
 \eqnapprox
   - 4 \mN
     \afrac{\Enu}{\mN + \Enu}^2 \ceta
     \non\\
 \eqnapprox
   - \afrac{4 \Enu^2}{\mN} \ceta
\~.
\label{eqn:dQQ_dOmega_eta}
\eeqn
\cheqn
 The minus signs
 appearing in Eq.~(\ref{eqn:dQQ_deta})
 and thus in Eq.~(\ref{eqn:dQQ_dOmega_eta})
 as well as
 in Eqs.~(\ref{eqn:dsigma_nuN_deta:SI})
 and (\ref{eqn:Q-dsigma_nuN_deta:SI})
 indicate that
 the recoil energy $Q$
 and the total cross section $\sigmanuNSM$
 decrease,
 while the recoil angle $\eta$ increases
 from 0 to $\pi / 2$
 (the equivalent recoil angle $\thetaNRnu$ decreases
  from $\pi / 2$ to 0).

\subsubsection{In the CM reference frame}
\label{subsubsec:CEnuNS:SM:QQ:psi}

 The nuclear recoil energy
 induced by \CEnuNS\
 can also be given exactly by
 the scattering angle
 in the center--of--momentum reference frame
 as
 \cite{Drukier84, Sahu20}
\beqn
     Q (\Enu, \psi)
 \=  \mN \cbrac{1 + 2 \afrac{\Enu}{\mN}^2 \bBig{1 - \cpsi}}^{1 / 2} - \mN
     \non\\
 \eqnapprox
     \frac{\Enu^2}{\mN} \bBig{1 - \cpsi}
\~.
\label{eqn:QQ_psi}
\eeqn
 This gives
\cheqna
\beqn
      \Dd{Q (\Enu, \psi)}{\psi}
 \eqnapprox
      \afrac{\Enu^2}{\mN} \spsi
\~,
\label{eqn:dQQ_dpsi}
\eeqn
 and
\cheqnb
\beqn
      \Dd{Q (\Enu, \psi)}{\Omega_{\psi}}
 \propto
      \frac{1}{\spsi} \cdot \Dd{Q (\Enu, \psi)}{\psi}
 \eqnapprox
      \frac{\Enu^2}{\mN}
  =   {\rm const.}
\~,
\label{eqn:dQQ_dOmega_psi}
\eeqn
\cheqn
 respectively.
 Eq.~(\ref{eqn:dQQ_dOmega_psi}) indicates that
 the angular distribution of
 the recoil energy
 is (approximately) isotropic
 in the center--of--momentum reference frame
 and inversely proportional to
 the mass of the target nucleus.

\subsection{Angular distribution of nuclear recoil flux}
\label{subsec:CEnuNS:SM:dsigma_deta}
\label{subsec:CEnuNS:SM:dsigma_dtheta}
\label{subsec:CEnuNS:SM:dsigma_dpsi}

 As the most important validation criterion
 in our Monte Carlo scattering--by--scattering simulation,
 we need
 the angular distribution of
 the nuclear recoil flux of
 3-D \CEnuNS.

\subsubsection[In the $\nuin$ coordinate system]
              {\boldmath
               In the $\nuin$ coordinate system}
\label{subsubsec:CEnuNS:SM:dsigma_deta}
\label{subsubsec:CEnuNS:SM:dsigma_dtheta}

 Combining Eq.~(\ref{eqn:dsigma_nuN_dQQ})
 with Eqs.~(\ref{eqn:QQ_eta}) and (\ref{eqn:dQQ_deta}),
 we can obtain
 the angular distribution of
 the nuclear recoil flux
 in the incoming--neutrino coordinate system
 as
\beqn
     \Dd{\sigmanuNSM}{\eta}
 \=  \Dd{\sigmanuNSM}{Q} \cdot \Dd{Q}{\eta}
     \non\\
 \eqnapprox
   - \afrac{G_F^2}{\pi} \abrac{2 \Enu^2}
     \cbigg{  \grmv^2            \sseta  \FSIQ
            + \grma^2 \bbrac{1 + \cceta} \FSDQ  }
     \sdeta
     \non\\
 \eqnapprox
   - \afrac{G_F^2}{\pi} \grmv^2 \abrac{4 \Enu^2} \FSIQ
     \ssseta \ceta
\~.
\label{eqn:dsigma_nuN_deta:SI}
\eeqn
 Multiplying
 the recoil energy $Q$,
 we obtain
 the angular distribution of
 the nuclear recoil energy
\beqn
     Q \aDd{\sigmanuNSM}{\eta}
 \approx
   - \afrac{G_F^2}{\pi} \grmv^2 \afrac{8 \Enu^4}{\mN} \FSIQ
     \ssseta \ccceta
\~.
\label{eqn:Q-dsigma_nuN_deta:SI}
\eeqn
 Additionally,
 it is easily to solve
 the recoil angles
 for the maximal recoil flux and energy
 as
\beq
     \eta
  =  60^{\circ}
\~,
     ~~~~ ~~~~ ~~~~ ~~~~ 
     {\rm and}
     ~~~~ ~~~~ ~~~~ ~~~~ 
     \eta
  =  45^{\circ}
\~,
\label{eqn:dsigma_nuN_dpsi:SI:eta:eta}
\label{eqn:Q-dsigma_nuN_dpsi:SI:eta:eta}
\eeq
 respectively.
 Or,
 equivalently,
 we have
\beq
     \thetaNRnu
  =  30^{\circ}
\~,
     ~~~~ ~~~~ ~~        
     {\rm and}
     ~~~~ ~~~~ ~~~~ ~~~~ 
     \thetaNRnu
  =  45^{\circ}
\~,
\label{eqn:dsigma_nuN_dpsi:SI:eta:theta}
\label{eqn:Q-dsigma_nuN_dpsi:SI:eta:theta}
\eeq
 respectively.

\subsubsection{In the CM reference frame}
\label{subsubsec:CEnuNS:SM:dsigma_dpsi}

 Combining Eq.~(\ref{eqn:dsigma_nuN_dQQ})
 with Eqs.~(\ref{eqn:QQ_psi}) and (\ref{eqn:dQQ_dpsi}),
 we can also obtain
 the angular distribution of
 the nuclear recoil flux and energy
 in the center--of--momentum reference frame
 as
 \cite{Drukier84}
\beqn
     \Dd{\sigmanuNSM}{\psi}
 \eqnapprox
     \afrac{G_F^2}{\pi} \afrac{\Enu^2}{2}
     \cbigg{  \grmv^2 \bBig{1 + \cpsi} \FSIQ
            + \grma^2 \bBig{3 - \cpsi} \FSDQ  }
     \spsi
     \non\\
 \eqnapprox
     \afrac{G_F^2}{\pi} \grmv^2 \afrac{\Enu^2}{2} \FSIQ \~
     \bBig{1 + \cpsi} \~ \spsi
\~,
\label{eqn:dsigma_nuN_dpsi:SI}
\eeqn
 and
\beqn
     Q \aDd{\sigmanuNSM}{\psi}
 \approx
     \afrac{G_F^2}{\pi} \grmv^2 \afrac{\Enu^4}{2 \mN} \FSIQ
     \ssspsi
\~,
\label{eqn:Q-dsigma_nuN_dpsi:SI}
\eeqn
 respectively.

\section{\boldmath
         Double--Monte Carlo scattering--by--scattering simulation of
         3-D \CEnuNS\ events}
\label{sec:3D-CEnuNS}

 In this section,
 we give first
 the overall workflow of
 the complete double--Monte Carlo scattering--by--scattering simulation process of
 3-dimensional coherent elastic neutrino--nucleus scattering.
 Then
 we describe
 the MC generation of
 the 3-D information
 (the energy,
  the incoming/scattering time
  and thus the moving direction)
 of incident Solar $\rmB$ neutrinos.
 In Sec.~\ref{subsec:3D-CEnuNS},
 we introduce
 the incoming--neutrino coordinate system
 and describe in detail
 the validation of
 each generated 3-D \CEnuNS\ event.

\subsection{Simulation workflow}
\label{subsec:workflow}
\begin{figure} [t!]
\begin{center}
 \includegraphics [width = 15.5 cm] {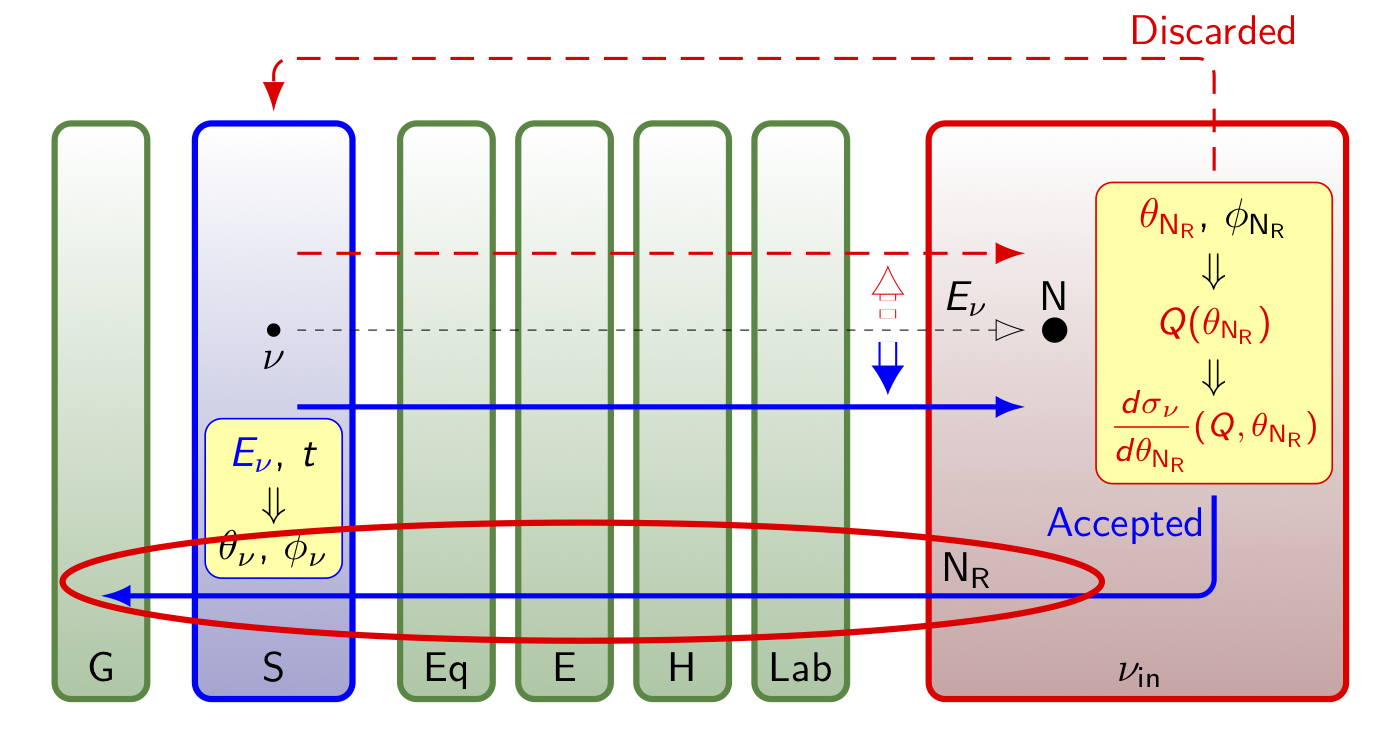}
\end{center}
\caption{
 The workflow of
 our double--Monte Carlo simulation and data analysis procedure of
 3-D \CEnuNS\ events.
 See the text
 for detailed descriptions.
}
\label{fig:workflow}
\end{figure}

 In this subsection,
 we describe
 the overall workflow of
 our double--Monte Carlo simulation and data analysis procedure of
 3-D \CEnuNS,
 sketched
 in Fig.~\ref{fig:workflow}
 in detail:

\begin{enumerate}
\item
 The 3-D information of
 incident Solar $\rmB$ neutrinos
 (the energy,
  the incoming/scattering time
  and thus the moving direction)
 is MC generated
 according to the theoretically estimated energy spectrum
 \cite{Solarnuspectrum}
 in the Ecliptic coordinate system
 (the blue subframe),
 which will be described
 in Sec.~\ref{subsec:Enu-S}.
\item
 The generated 3-D neutrino energy/momentum
 will be transformed through
 the laboratory--independent
 (Equatorial
  and Earth)
 coordinate systems
 as well as
 the laboratory--dependent
 (horizontal
  and laboratory)
 coordinate systems
 (the green subframes,
  see Appendix \ref{appx:XYZ})
 and
 at the end
 into the incoming--neutrino coordinate system
 (the red subframe,
  defined in Sec.~\ref{subsubsec:XYZ_nu}).
\item
 In the incoming--neutrino coordinate system,
 the 3-D \CEnuNS\ process
 will also be MC simulated
 by generating
 an equivalent recoil angle $\thetaNRnu$
 and an orientation of the scattering plane $\phiNRnu$
 (defined in Sec.~\ref{subsubsec:phi_theta_NR_nu}).
 They define the recoil direction of
 the scattered target nucleus
 and the former,
 combined with the neutrino incident energy $\Enu$,
 will then be used for estimating
 the transferred recoil energy to
 the scattered target nucleus,
 $Q \~ (\Enu, \eta = \pi / 2 - \thetaNRnu)$
 and
 to validate
 the differential \CEnuNS\ cross section
 with respect to
 the recoil angle
 $\eta$,
 $d\sigmanuNSM / d\eta \~ (Q, \eta)$,
 given by Eq.~(\ref{eqn:dsigma_nuN_deta:SI})
 as our event validation criterion
 (see Sec.~\ref{subsec:CEnuNS:SM:dsigma_deta}
  for details).
\item
 The MC generated equivalent recoil angle $\thetaNRnu$
 and orientation of the scattering plane $\phiNRnu$ of
 the accepted recoil events
 will be transformed (back)
 through all celestial coordinate systems
 (indicated by the lower solid blue arrow).
 All these 3-D information of
 the scattered target nucleus
 accompanied with
 the corresponding recoil energy $Q \~ (\Enu, \eta)$
 as well as
 that of the scattering neutrino
 in different coordinate systems
 (the upper solid blue arrow)
 will be recorded
 for further analyses.
\item
 For the invalid cases,
 in which
 the estimated recoil energies
 are either out of the experimental measurable energy window
 or suppressed by the validation criterion
 $d \sigmanuNSM / d \eta \~ (Q, \eta)$,
 the generated 3-D information
 on the incident neutrinos
 (the lower dashed red arrow)
 (and that on the recoiled nucleus)
 will be discarded
 and the generation/validation process of
 one \CEnuNS\ event
 will be restarted from the Ecliptic coordinate system
 (the upper dashed red arrow).
\end{enumerate}
\subsection[MC generation of Solar $\rmB$ neutrinos
            in the Ecliptic coordinate system]
           {\boldmath
            MC generation of Solar $\rmB$ neutrinos
            in the Ecliptic coordinate system}
\label{subsec:Enu-S}

 In this subsection,
 we describe
 the MC generation process of
 the 3-dimensional information of
 incident Solar $\rmB$ neutrinos
 in our directional direct Dark Matter detectors
 (the energy,
  the incoming/scattering time
  and thus the incident direction)
 in the Ecliptic coordinate system.

\subsubsection[Spectrum of incident Solar $\rmB$ neutrinos]
              {\boldmath
               Spectrum of incident Solar $\rmB$ neutrinos}
\label{subsubsec:dPhidE}

 For generating
 the energy of
 incident Solar $\rmB$ neutrinos
 in the Ecliptic coordinate system,
 we use the public numerical data of
 the theoretically estimated energy spectrum
 provided on the {\it ``Software and Data for Solar Neutrino Research''} website
 \cite{Solarnuspectrum}.

\begin{figure} [t!]
\begin{center}
\includegraphics [width = 13 cm] {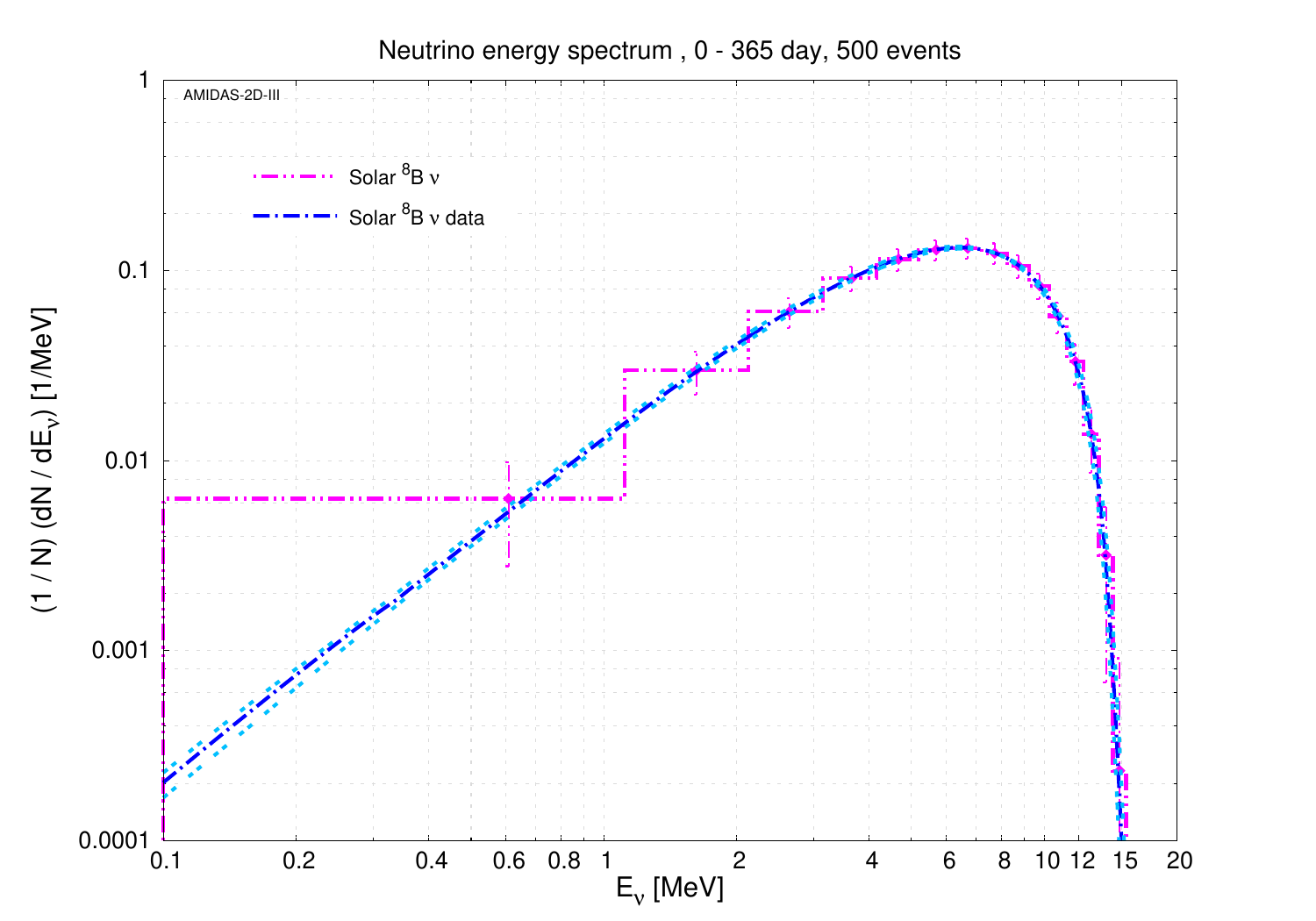}
\end{center}
\caption{
 The energy spectrum of
 the incident Solar $\rmB$ neutrinos
 generated by using the public numerical data of
 the theoretically estimated energy spectrum
 provided on the {\it ``Software and Data for Solar Neutrino Research''} website
 \cite{Solarnuspectrum}.
 500 total events on average
 (Poisson--distributed)
 in one experiment
 in one entire year
 have been generated.
 The dash--dotted blue curve
 indicates the theoretically estimated energy spectrum of
 Solar $\rmB$ neutrinos,
 while
 the dash--double--dotted magenta histogram
 and the thin vertical dash--double--dotted magenta lines
 show
 the (1$\sigma$ Poisson statistical uncertainties on the) number of
 the generated neutrino energies.
 See the text for further details.
}
\label{fig:nu_N_E-B8-05000-00000}
\end{figure}

 In Fig.~\ref{fig:nu_N_E-B8-05000-00000},
 we show
 the generated energy spectrum of
 incident Solar $\rmB$ neutrinos.
 500 total events on average
 (Poisson--distributed)
 in one experiment
 in one entire year (0 to 365 day)
 have been generated
 and
 5,000 experiments have been simulated.
 The dash--dotted blue curve
 indicates the theoretically estimated energy spectrum of
 Solar $\rmB$ neutrinos
 and
 the double--dashed cyan curves
 indicate the 3$\,\sigma$ upper and lower bounds of
 the estimated energy spectrum,
 while
 the dash--double--dotted magenta histogram
 and the thin vertical dash--double--dotted magenta lines
 show
 the (1$\sigma$ Poisson statistical uncertainties on the) number of
 the generated neutrino energies.

\subsubsection[Angular distribution of incident Solar $\rmB$ neutrinos]
              {\boldmath
               Angular distribution of incident Solar $\rmB$ neutrinos}
\label{subsubsec:N_ang-Eq}

 Since
 the distance between the Sun and the Earth
 ($1.496 \times 10^8$ km
  \cite{RPP24AP})
 is much larger than
 the Solar equatorial radius
 ($6.957 \times 10^5$ km
  \cite{RPP24AP}),
 in our simulations
 presented in this paper,
 we assumed that
 all Solar neutrinos
 move from the Solar center to the Earth's center,
 namely,
 in the Ecliptic plane.
 This means that,
 in the Ecliptic coordinate system,
 the elevation of incident Solar $\rmB$ neutrinos
 is set to be zero
\beq
     \thetanuS
  =  0
\~.
\label{eqn:thetanuS}
\eeq
 On the other hand,
 given an incoming/scattering time $t$,
 one can determine
 the azimuthal angle of incident Solar $\rmB$ neutrinos
 in the Ecliptic coordinate system
 by%
\footnote{
 In our simulation package,
 firstly,
 the primary direction (the $\xS$--axis) of
 the Ecliptic coordinate system
 is the direction
 pointing from the Solar center to the Earth's center
 at 12 o'clock midnight (the end) of
 the date of
 the 79th day (the March 20th).
 Secondly,
 the Earth's orbit around the Sun
 is assumed to be perfectly circular
 in the Ecliptic plane
 and the Earth's orbital speed
 is thus a constant.
}
\beq
     \phinuS
  =  2 \pi \afrac{t - 79.0}{365.0}
\~.
\label{eqn:phinuS}
\eeq
 \def \nuSource     {B8}
 \def \ShortFrame   {G}
 \def \EventNumber  {05000}
\begin{figure} [t!]
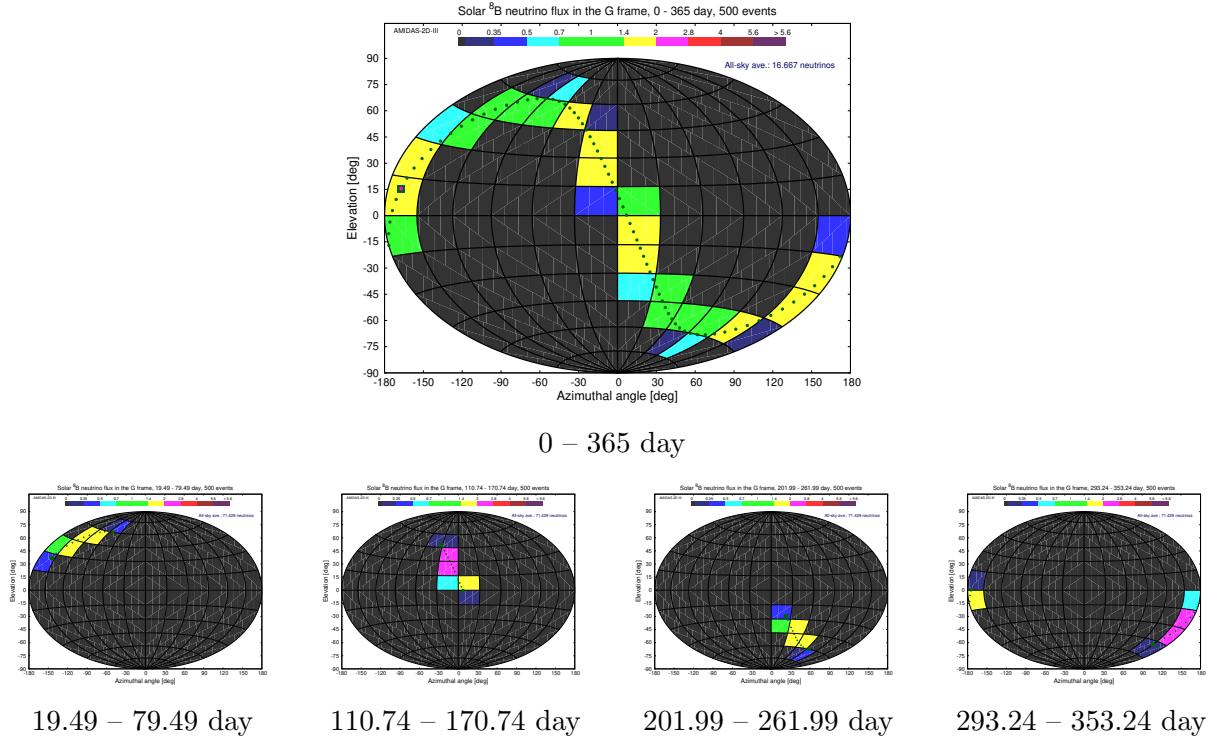

\begin{center}
 \InsertSubfigureNangAnnual
  {nu_N}
  {\nuSource}
  {}
\end{center}
\caption{
 The angulardistributions of
 incident Solar $\rmB$ neutrinos
 in the Galactic coordinate system.
 500 total events on average
 (Poisson--distributed)
 in one experiment
 in one entire year
 and four {\em advanced} seasons of 60 days each
 (see Table \ref{tab:period_year})
 have been generated
 and binned into \mbox{12 $\times$ 12} bins
 for the elevation and the azimuthal angle,
 respectively.
 The dark--green square and the (pink) disk
 indicate the starting and the end points of
 the direction pointing
 from the Solar center to the Earth's center
 in each observation period.
 The horizontal color bar on the top of each plot
 indicates
 the mean value of the recorded event number
 (averaged over all simulated experiments)
 in each angular bin
 in unit of the all--sky average value
 (500 events/30 or 7 non--empty bins
  $=$ 16.67 or 71.43 events/bin
  here).
}
\label{fig:nu_N_ang-B8-G-05000}
\end{figure}
 \def \ShortFrame   {Eq}
 \def \LabName      {SNOLAB}
 \def \LabLocation  {(46.47$^{\circ}$N, 81.19$^{\circ}$W)}
\begin{figure} [t!]
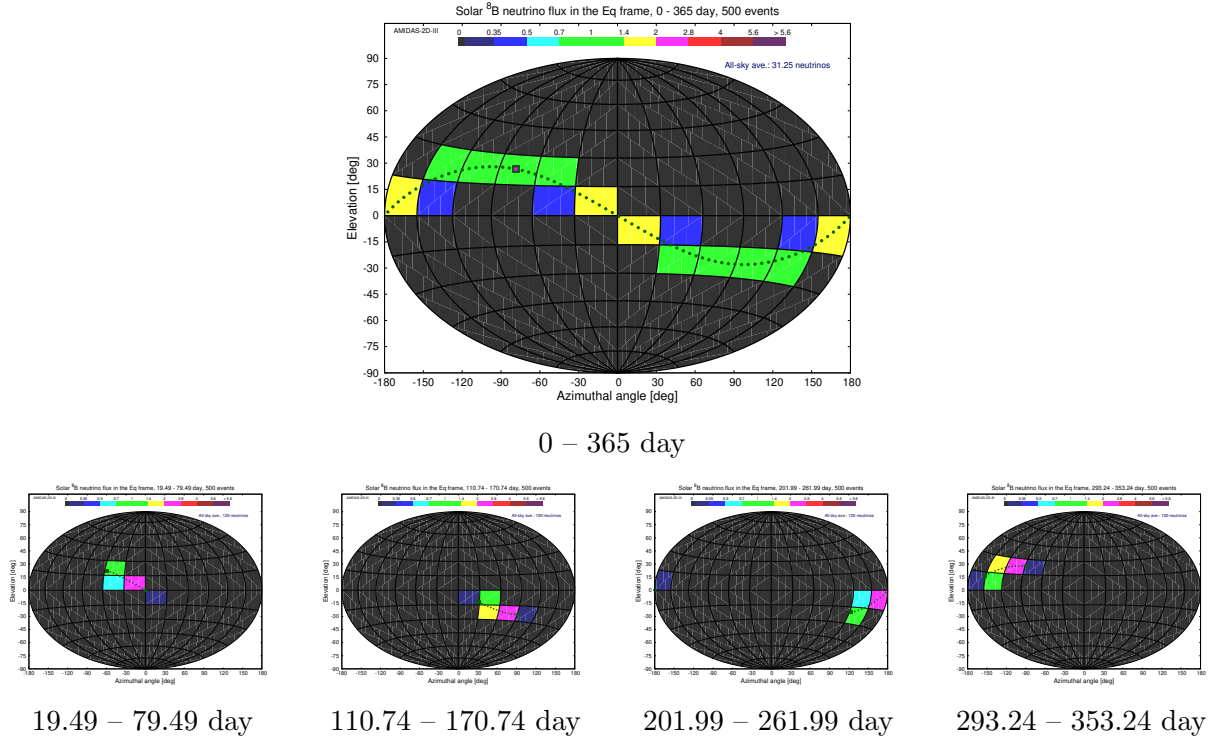

\begin{center}
 \InsertSubfigureNangAnnual
  {nu_N}
  {\nuSource}
  {}
\end{center}
\caption{
 The same as Figs.~\ref{fig:nu_N_ang-B8-G-05000},
 except that
 the angular distributions are
 in the Equatorial coordinate system.
 The all--sky average value here are
 500 events/16 or 4 or 5 non--empty bins
 $=$ 31.25 or 125 or 100 events/bin,
 respectively.
}
\label{fig:nu_N_ang-B8-Eq-05000}
\end{figure}
 \def \ShortFrame   {Lab}
 \def \LabName      {SNOLAB}
 \def \LabLocation  {(46.47$^{\circ}$N, 81.19$^{\circ}$W)}
\begin{figure} [t!]
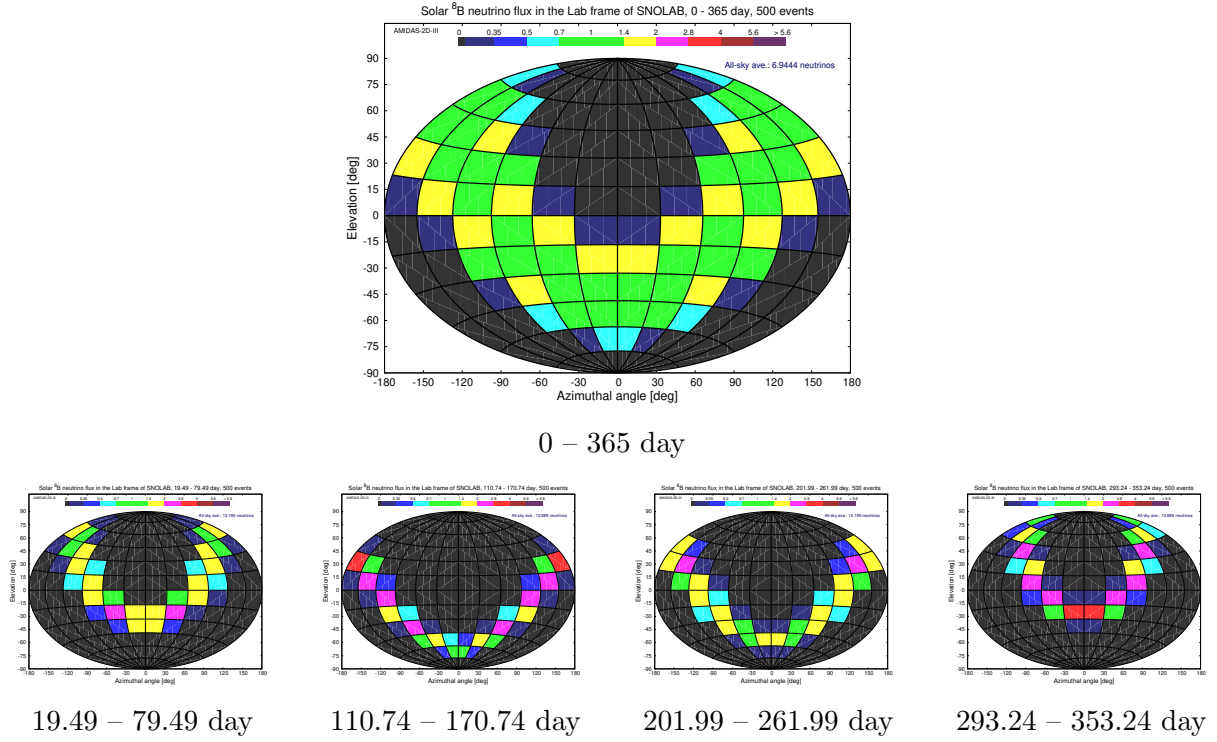

\begin{center}
 \InsertSubfigureNangAnnual
  {nu_N}
  {\nuSource}
  {-\LabName}
\end{center}
\caption{
 The same as Figs.~\ref{fig:nu_N_ang-B8-G-05000}
 and               \ref{fig:nu_N_ang-B8-Eq-05000},
 except that
 the angular distributions are
 in the laboratory coordinate system of
 \LabName\
 \LabLocation.
 The all--sky average value here are
 500 events/72 or 41 or 36 non--empty bins
 $=$ 6.94 or 12.20 or 13.89 events/bin,
 respectively.
}
\label{fig:nu_N_ang-B8-Lab-05000-\LabName}
\end{figure}

 In Figs.~\ref{fig:nu_N_ang-B8-G-05000}
 to \ref{fig:nu_N_ang-B8-Lab-05000-\LabName},
 we show
 the angular distributions of
 incident Solar $\rmB$ neutrinos
 in the Galactic, Equatorial, and laboratory coordinate systems,
 respectively.
 500 total events on average
 (Poisson--distributed)
 in one experiment
 in one entire year
 and four {\em advanced} seasons of 60 days each
 (see Table \ref{tab:period_year})
 have been generated
 and binned into 12 $\times$ 12 bins
 for the elevation and the azimuthal angle,
 respectively.
 5,000 experiments have been simulated.
 The dark--green square and the (pink) disk
 indicate the starting and the end points of
 the direction pointing
 from the Solar center to the Earth's center
 in each observation period
 (see also Fig.~\ref{fig:E-Eq-S}).
 The horizontal color bar on the top of each plot
 indicates
 the mean value of the recorded event number
 (averaged over all simulated experiments)
 in each angular bin
 in unit of the all--sky average value.

\subsubsection[Incoming/scattering time
               and observation periods of
               3-D \CEnuNS\ events]
              {\boldmath
               Incoming/scattering time
               and observation periods of
               3-D \CEnuNS\ events}
\label{subsubsec:N_t}
\begin{table} [b!]
\small
\begin{center}
\renewcommand{\arraystretch}{1.5}
\begin{tabular}{|| c || c | c ||}
\hline
\hline
 \makebox[5 cm][c]{Option}             &
 \makebox[5 cm][c]{Central date (day)} &
 \makebox[5 cm][c]{Period       (day)} \\
\hline
\hline
 One entire year
 & ---    & \PeriodAa \\
\hline
 \multirow{4}{*}{Four advanced seasons}
 &  49.49 & \PeriodCa \\
 & 140.74 & \PeriodCb \\
 & 231.99 & \PeriodCc \\
 & 323.24 & \PeriodCd \\
\hline
\hline
\end{tabular}
\end{center}
\caption{
 Two options
 for the observation periods
 in a 365-day year
 considered
 in our simulations
 for demonstrating
 the annual variations of
 the angular distributions of
 the recoil flux and energy of
 the scattered target nuclei.
}
\label{tab:period_year}
\end{table}

 Since,
 in the Ecliptic point of view,
 \CEnuNS\ events
 should be observed randomly and constantly,
 in our simulations,
 we considered a constant probability
 for generating the
 UTC (Coordinated Universal Time)
 incoming/scattering time of
 the recorded \CEnuNS\ signals:
\beq
     f_{t}(t)
  =  1
\~,
     ~~~~ ~~~~ ~~ 
     t \in [t_{\rm start},~t_{\rm end}]
\~.
\label{eqn:N_t_S}
\eeq
 Moreover,
 as listed in Table \ref{tab:period_year},
 except for
 one entire year,
 for demonstrating
 the annual variations of
 the angular distributions of
 the recoil flux and energy of
 the scattered target nuclei
 (as well as
  for comparing with our earlier work on
  WIMP--nucleus scattering signals),
 we have considered
 four advanced seasons
 with the central dates on
 the February 19th     (49.49 day),
 the May 21st          (140.74 day),
 the August 20th       (231.99 day),
 and the November 20th (323.24 day),
 respectively.
 This is because
 the relative velocity of the Earth
 to the Galactic Dark Matter halo
 should be the maximum (minimum),
 when its orbital velocity is (anti--)parallel to
 the projection of the direction of the Solar movement
 on the Ecliptic plane
 around the 21st of May
 (the 20th of November)
 \cite{DMDDD-N}.
 In the four--season option,
 we considered
 a 60-day observation period.

\subsection[MC generation of 3-D \CEnuNS\ events
            in the incoming--neutrino coordinate system]
           {\boldmath
            MC generation of 3-D \CEnuNS\ events
            in the incoming--neutrino coordinate system}
\label{subsec:3D-CEnuNS}

 As described in Sec.~\ref{subsec:workflow},
 each generated 3-D moving direction/momentum of
 incident Solar $\rmB$ neutrino
 will be transformed through
 different celestial coordinate systems
 and,
 at the end,
 into
 the incoming--neutrino $\nuin$ coordinate system.
 In this subsection,
 we focus then on
 the core part of
 our simulation procedure:
 the generation and the validation of
 3-D \CEnuNS\ events
 in the incoming--neutrino coordinate system.

 We give first
 our definition of
 the incoming--neutrino coordinate system
 as well as
 the definitions of
 the orientation of the scattering plane
 and the equivalent recoil angle.
 Then
 we discuss
 the validation process of
 our MC simulation
 by taking into account
 the cross section suppression
 on each generated recoil energy
 in detail.

\subsubsection{Definition of the incoming--neutrino coordinate system}
\label{subsubsec:XYZ_nu}

 In Fig.~\ref{fig:nu-Lab},
 we sketch
 the definition of
 the (light--green) incoming--neutrino coordinate system
 in the (dark--green) laboratory coordinate system.
 The $\znu$--axis is defined as usual as
 the direction of the momentum of
 the incoming Solar $\rmB$ neutrino of interest
 $\bfpnu$.
 $\thetanuLab$ and $\phinuLab$ indicate
 the elevation and the azimuthal angle of
 $\bfpnu$
 measured in the laboratory coordinate system,
 respectively.
 The $\xnu$--axis is perpendicular to the $\znu$--axis
 and lies in the $\zLab$--$\znu$ plane.
 Then
 the $\ynu$--axis is defined
 by the right--handed convention.
 Note that,
 firstly,
 the $\ynu$--axis lies always in the $\xLab$--$\yLab$ plane,
 since
 it is by definition
 perpendicular to the $\xnu$--$\zLab$--$\znu$ plane.
 Secondly,
 the center of the incoming--neutrino coordinate system
 is at the position of the scattered target nucleus
 before scattering
 (see also Fig.~\ref{fig:NR-nu-Lab}).

\InsertSKPPlotS
 {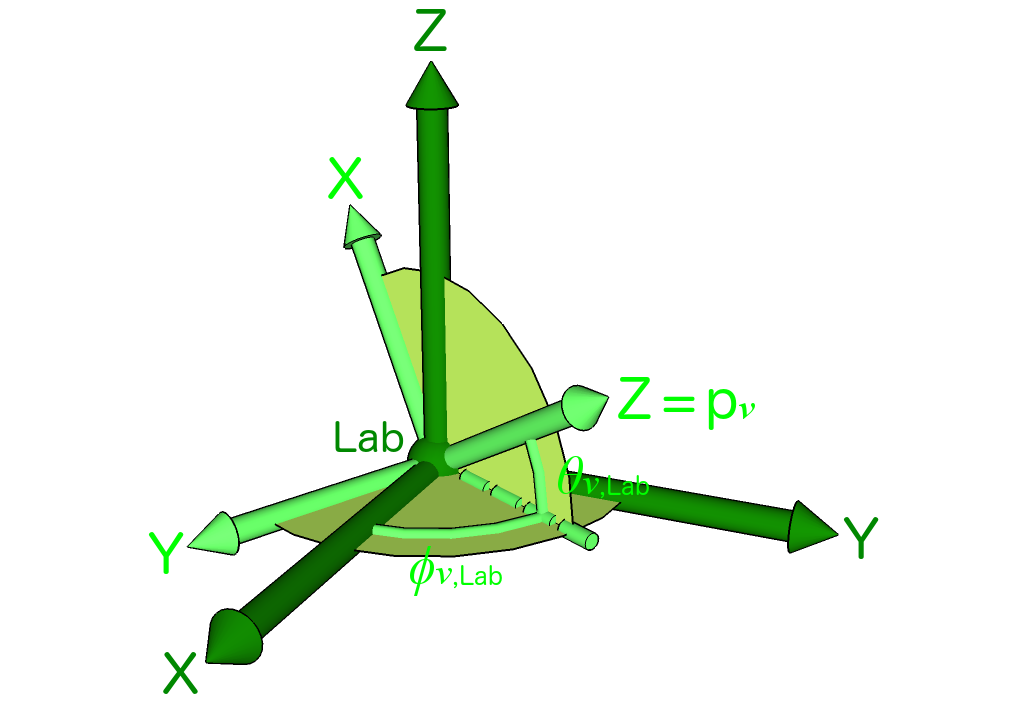}
 {The definition of
  the (light--green) incoming--neutrino coordinate system
  in the (dark--green) laboratory coordinate system.
  The $\znu$--axis is defined as usual as
  the direction of the momentum of
  the incoming Solar $\rmB$ neutrino of interest
  $\bfpnu$.
  $\thetanuLab$ and $\phinuLab$ indicate
  the elevation and the azimuthal angle of
  $\bfpnu$
  measured in the laboratory coordinate system,
  respectively.
  The $\xnu$--axis is perpendicular to the $\znu$--axis
  and lies in the $\zLab$--$\znu$ plane.
  Then
  the $\ynu$--axis is defined
  by the right--handed convention.
  }
 \subsubsection{Generation of nuclear recoil directions}
 \label{subsubsec:phi_theta_NR_nu}

 In Fig.~\ref{fig:NR-nu-Lab},
 we sketch
 the process of
 one single 3-D \CEnuNS\ event:
 $\nu_{\rm in/out}$ indicate
 the incoming and the outgoing directions of
 the scattering Solar $\rmB$ neutrino,
 respectively.
 While
 $\zeta$ indicates
 the scattering angle of
 the outgoing neutrino $\nu_{\rm out}$
 (measured from the $\znu$--axis),
 $\eta$ is
 the recoil angle of
 the scattered target nucleus $\rm N_R$.

 According to
 our definition of the incoming--neutrino coordinate system,
 the orientation of
 the ($\bfpnu'$--$\znu$--$\bfpN'$) scattering plane of
 this single scattering event
 can be specified by
 the azimuthal angle of
 the recoil direction of
 the scattered target nucleus,
 $\phiNRnu$,
 which
 should be azimuthal symmetric
 around the $\znu$--axis
 and is thus
 generated with a constant probability
 in our simulation package:
\beq
     f_{{\rm N_R}, \nuin, \phi} (\phiNRnu)
  =  1
\~,
     ~~~~ ~~~~ ~~ 
     \phiNRnu \in (-\pi,~\pi]
\~.
\label{eqn:f_NR_phiNRnu}
\eeq
 On the other hand,
 we use
 the equivalent recoil angle
\beq
      \thetaNRnu
 \in  [0,~\pi / 2]
\~,
\label{eqn:thetaNRnu_range}
\eeq
 and
 the expression for
 $d\sigmanuNSM / d\thetaNRnu \~ (Q, \thetaNRnu)$
 modified from Eq.~(\ref{eqn:dsigma_nuN_deta:SI})
 as the generating probability distribution:
\beqn
 \conti
     f_{{\rm N_R}, \nuin, \theta} (\thetaNRnu)
     \non\\
 \=  \Dd{\sigmanuNSM}{\thetaNRnu}
     \non\\
 \eqnapprox
     \afrac{G_F^2}{\pi} \abrac{2 \Enu^2}
     \cbigg{  \grmv^2            \cctheta  \FSIQ
            + \grma^2 \bbrac{1 + \sstheta} \FSDQ  }
     \sdtheta
     \non\\
 \eqnapprox
     \afrac{G_F^2}{\pi} \grmv^2 \abrac{4 \Enu^2} \FSIQ
     \stheta \ccctheta
\~.
\label{eqn:f_NR_thetaNRnu}
\eeqn
 Remind here that
 the recoil energy $Q$
 depends on both of
 the energy of the incident Solar $\rmB$ neutrino
 as well as
 the equivalent recoil angle $\thetaNRnu$.

\InsertSKPPlotS
 {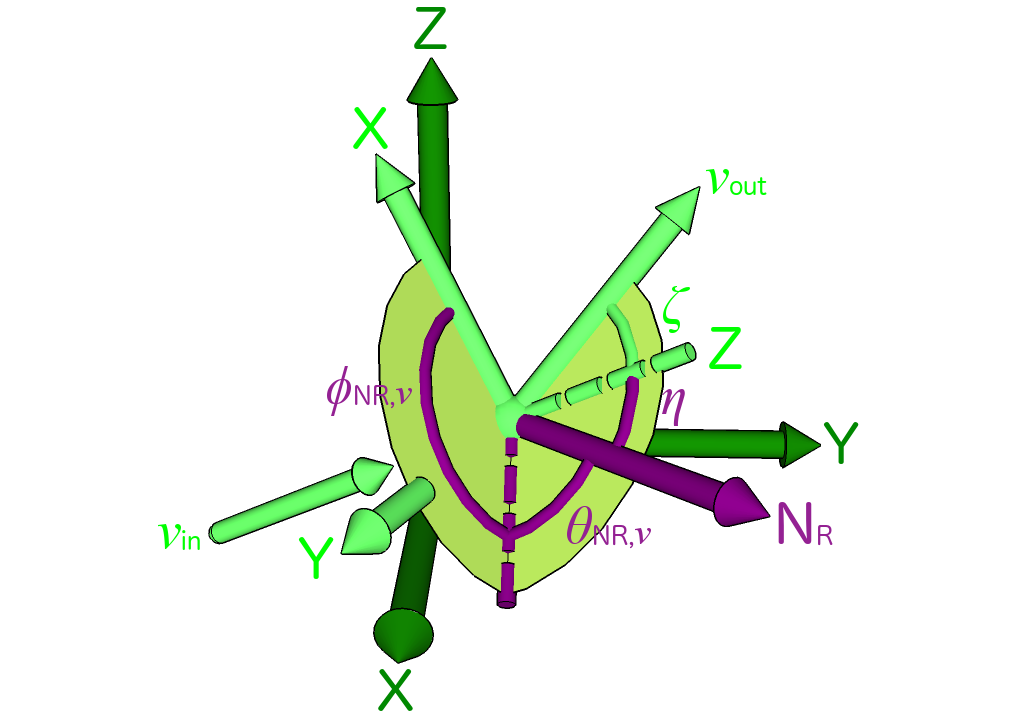}
 {A 3-D \CEnuNS\ event
  in the (light--green) incoming--neutrino
  and the (dark--green) laboratory coordinate systems.
  $\zeta$ and $\eta$ are
  the scattering angle of
  the outgoing neutrino $\nu_{\rm out}$
  and
  the recoil angle of
  the scattered target nucleus $\rm N_R$
  measured in the incoming--neutrino coordinate system of
  this single scattering event,
  respectively.
  (See also Fig.~\ref{fig:nu-N}.)
  The azimuthal angle of
  the recoil direction of $\rm N_R$
  in this incoming--neutrino coordinate system,
  $\phiNRnu$,
  indicates the orientation of the scattering plane,
  whereas
  the elevation of
  the recoil direction of $\rm N_R$,
  $\thetaNRnu$,
  is namely the complementary angle of
  the recoil angle
  $\eta$.
  }
\subsubsection{Nuclear form factors}
\label{subsubsec:FQ_SI/SD}

 In our simulation package,
 we adopt
 the commonly used analytic form
 for the elastic nuclear form factor
 \cite{SUSYDM96}:
\beq
     F_{\rm SI}^2 (Q)
  =  \bfrac{3 j_1(q R_1)}{q R_1}^2 e^{-(q s)^2}
\~,
\label{eqn:FQ_SI_WS}
\eeq
 as well as
 the thin--shell form factor
 \cite{Lewin96}:
\beq
     F_{\rm SD}^2 (Q)
  =  \cleft{\renewcommand{\arraystretch}{1.5}
            \begin{array}{l l l}
             j_0^2(q R_1)                      \~, & ~~~~ ~~~~ & 
             {\rm for}~q R_1 \le 2.55~{\rm or}~q R_1 \ge 4.5 \~, \\
             {\rm const.} \simeq 0.047         \~, &           &
             {\rm for}~2.5 5 \le q R_1 \le 4.5 \~,
            \end{array}}
\label{eqn:FQ_SD_TS}
\eeq
 for the SI vector
 and the SD axial--vector%
\footnote{
 This may theoretically not be realistic,
 but acceptable for
 a first--step approximate demonstration of
 the effects of
 non--negligible SD contribution
 combined with
 the form factor suppression.
}
 neutrino--nucleus cross sections,
 respectively.
 Here
 $j_1(x)$ and $j_0(x)$ are the spherical Bessel functions,
\beq
     q
  =  \sqrt{2 \mN Q}
\~,
\label{eqn:qq}
\eeq
 is the transferred 3-momentum,
 and
 for the effective nuclear radius,
 we use
\beq
     R_1
  =  \sqrt{R_A^2 - 5 s^2}
\~,
\label{eqn:R1}
\eeq
 with
\beq
     R_A
 \simeq
     1.2 \~ A^{1 / 3} \~ {\rm fm}
\~,
\label{eqn:RA}
\eeq
 and a nuclear skin thickness
\beq
     s
 \simeq
     1 \~ {\rm fm}
\~.
\label{eqn:ss}
\eeq
\begin{figure} [t!]
\begin{center}
 \includegraphics [width = 13 cm] {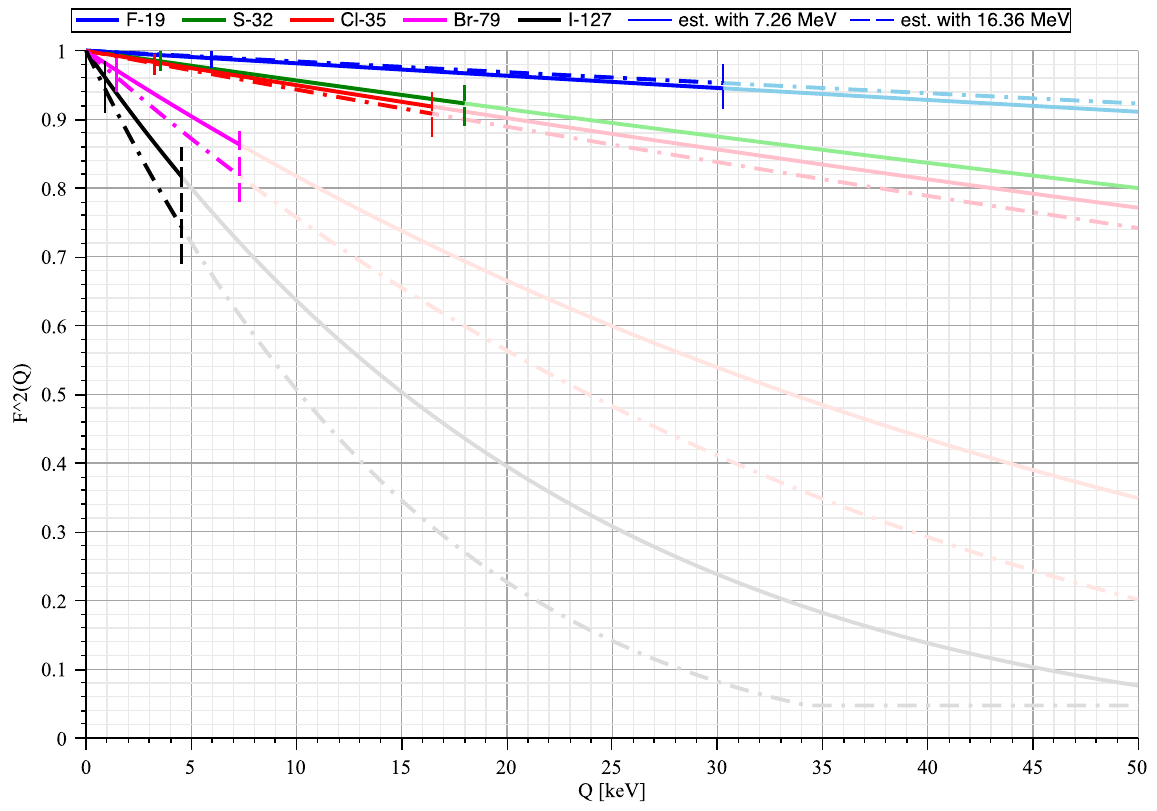}
\end{center}
\caption{
 Nuclear form factors of
 the $\rmF$     (blue),
 the $\rmS$     (green),
 the $\rmCl$    (red),
 the $\rmBr$    (magenta),
 and the $\rmI$ (black) nuclei
 as functions of the recoil energy
 up to 50 keV.
 The solid and dash--dotted curves
 indicate the form factors
 corresponding to the SI and SD cross sections,
 $F_{\rm SI}^2 (Q)$ and $F_{\rm SD}^2 (Q)$,
 given in Eqs.~(\ref{eqn:FQ_SI_WS}) and (\ref{eqn:FQ_SD_TS}),
 respectively.
 The thin vertical dashed cut--off lines
 indicate the maximal transferable recoil energies
 estimated by Eq.~(\ref{eqn:Qmax_Enu})
 with the maximal estimated energy of
 Solar $\rmB$ neutrinos of 16.36 MeV,
 while
 the short solid lines
 on the top--left corner
 are estimated
 with the root--mean--square energy
 $E_{\nu, {\rm rms}} = 7.26~{\rm MeV}$
 of Solar $\rmB$ neutrinos.
}
\label{fig:FQ}
\end{figure}

 In Fig.~\ref{fig:FQ},
 we show
 the recoil--energy dependence of
 the nuclear form factors corresponding to
 the SI (solid) and SD (dash--dotted) cross sections,
 $F_{\rm SI}^2 (Q)$ and $F_{\rm SD}^2 (Q)$,
 given in Eqs.~(\ref{eqn:FQ_SI_WS}) and (\ref{eqn:FQ_SD_TS}),
 respectively.
 Five frequently used target nuclei
 have been considered:
 $\rmF$     (blue),
 $\rmS$     (green),
 $\rmCl$    (red),
 $\rmBr$    (magenta),
 and $\rmI$ (black).
 The thin vertical dashed cut--off lines
 indicate the maximal transferable recoil energies
 estimated by Eq.~(\ref{eqn:Qmax_Enu})
 with the maximal estimated energy of
 Solar $\rmB$ neutrinos of 16.36 MeV,
 while
 the short solid lines
 on the top--left corner
 are estimated
 with the root--mean--square energy
 $E_{\nu, {\rm rms}} = 7.26~{\rm MeV}$
 of Solar $\rmB$ neutrinos.

\section{Numerical results}
\label{sec:NR}

 In this section,
 we demonstrate
 the angular distributions of
 the nuclear recoil flux
 and the (accumulated and average) recoil energy
 scattered by incident Solar $\rmB$ neutrinos
 and
 observed in the incoming--neutrino,
 the laboratory/location--dependent laboratory
 as well as
 the laboratory/location--independent Equatorial and Galactic
 coordinate systems,
 respectively.

 Two spin--sensitive nuclei
 used in directional direct DM/WIMP detection experiments:
 $\rmF$ and $\rmI$
 have been considered
 as our targets.
 We simulate
 5,000 experiments
 with 500 accepted events
 (blue arrows in Fig.~\ref{fig:workflow})
 on average
 (Poisson--distributed)%
\footnote{
 Note that,
 in our numerical simulation package,
 the actual number of accepted events
 in each simulated experiment
 is Poisson--distributed around
 the expectation value.
}
 in one observation period
 (\mbox{365 days/year}
  or 60 days/season)
 for one laboratory/target nucleus.
 Note that
 we assume simply that,
 firstly,
 the experimental threshold energies
 for all considered target nuclei
 are negligible;
 secondly,
 all experimental systematic uncertainties
 as well as
 the uncertainty on the measurement of the recoil energy
 could be ignored.

 As a comparison,
 we review briefly
 the angular distributions of
 the nuclear recoil flux/energy
 induced by Galactic halo WIMPs.
 The interested reader
 can also refer to Ref.~\cite{DMDDD-NR}
 for more simulation results and detailed discussions
 with different WIMP masses
 and different target nuclei.

\subsection{Nuclear recoil spectrum}
\label{subsec:dRdQ}

 In this subsection,
 we discuss first
 the nuclear recoil spectrum of
 3-D coherent elastic neutrino--nucleus scattering events,
 which can be theoretically estimated by
 \cite{Strigari09a,
       Guetlein10,
       Billard13b, Ruppin14,
       OHare16,
       Papoulias18,
       Abdullah20,
       Gaspert21}
\beq
     \Dd{\RnuN}{Q}
  =  \frac{1}{\mN}
     \int_{\Enumin (Q)}^{\Enumax}
     \adPhinudEnu \aDd{\sigmanuN}{Q}
     d\Enu
\~.
\label{eqn:dRnuNdQ}
\eeq
 Here
 $d\Phi_{\nu} / d\Enu$ is
 the neutrino spectrum
 shown in Fig.~\ref{fig:dPhidEnu},
 $d\sigmanuN / dQ$ is
 the differential \CEnuNS\ cross section,
 $\Enumax$ is
 the maximal energy of
 incident (Solar) neutrinos,
 and
 the factor $1 / \mN$ gives
 the number of target nuclei
 per unit mass of detector materials.
 By solving Eq.~(\ref{eqn:Qmax_Enu}),
 we can obtain
 the minimal--required neutrino incident energy
 for transferring nuclear recoil energy $Q$
 as
 \cite{Harnik12}
\beqn
     \Enumin (Q)
  =  \frac{Q + \sqrt{Q (Q + 2 \mN)}}{2}
 \eqnapprox
     \sfrac{\mN Q}{2}
\~.
\label{eqn:Enumin(Q)}
\eeqn
\begin{figure} [t!]
\begin{center}
 \begin{subfigure} [c] {13 cm}
  \includegraphics [width = 13 cm] {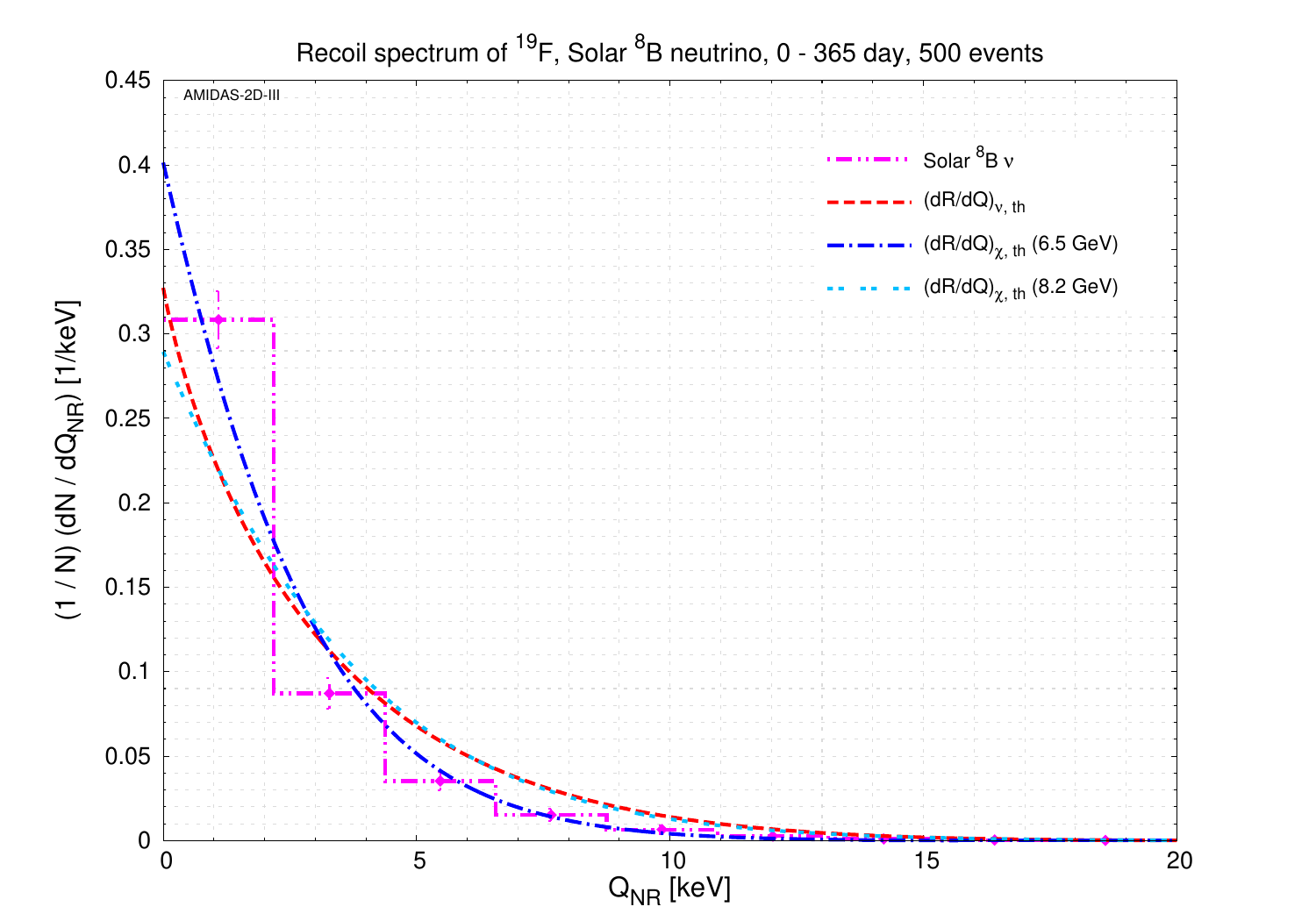}
 \caption{Off $\rmF$}
 \end{subfigure}
 \\
 \vspace{0.25 cm}
 \begin{subfigure} [c] {13 cm}
  \includegraphics [width = 13 cm] {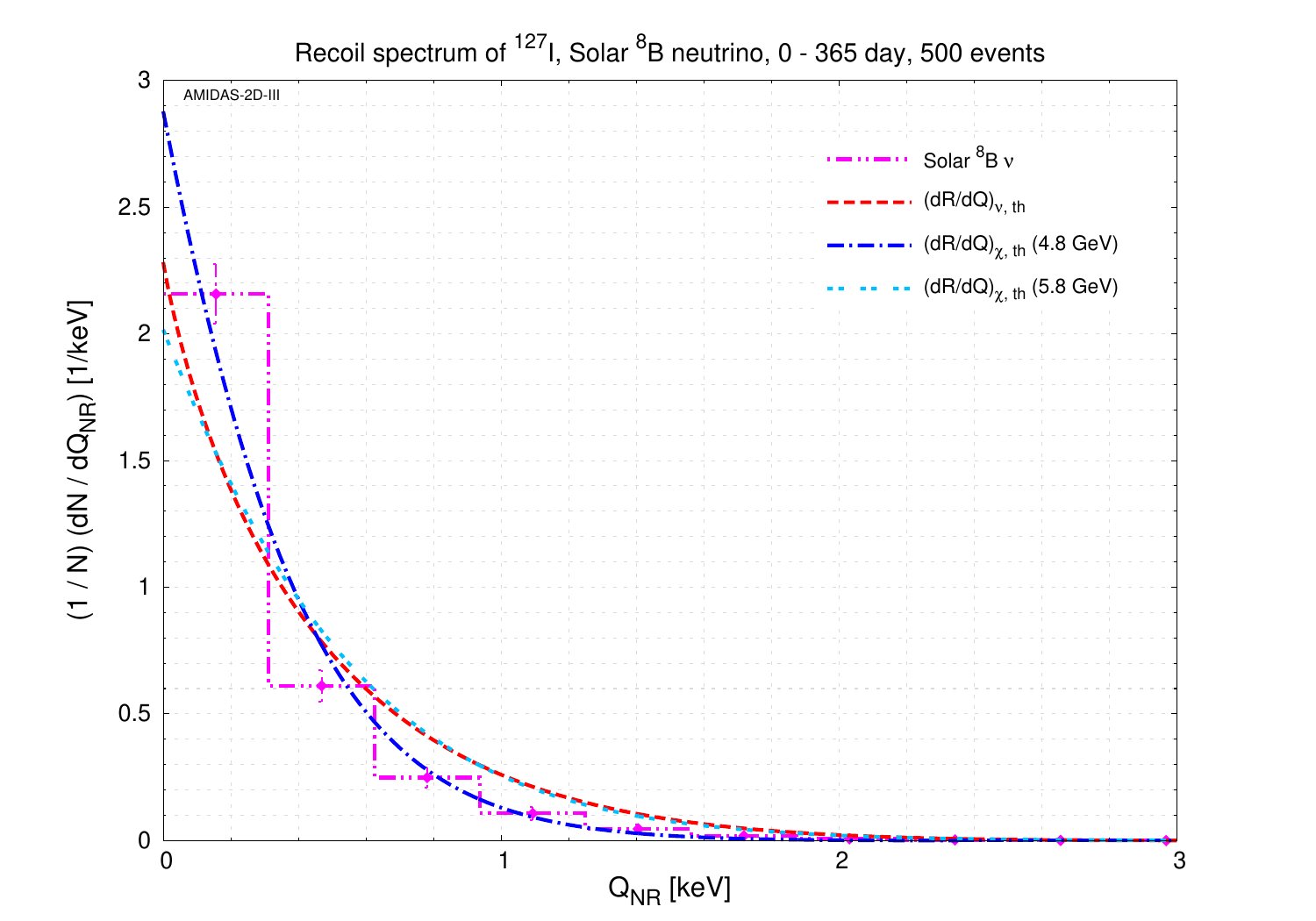}
 \caption{Off $\rmI$}
 \end{subfigure}
\end{center}
\caption{
 Nuclear recoil spectra of \CEnuNS\ off
 (a) $\rmF$ and (b) $\rmI$ target nuclei
 induced by Solar $\rmB$ neutrinos.
 500 accepted \CEnuNS\ events on average
 in one entire year
 have been recorded.
 While
 the dashed red curve
 indicates the theoretical \CEnuNS\ spectrum
 estimated by Eq.~\ref{eqn:dsigma_nuN_dQQ},
 the dash--dotted blue and double--dashed cyan curves
 are the theoretical WIMP--nucleus scattering spectra
 estimated by Eq.~\ref{eqn:dRdQ_SISD}
 with a WIMP mass of (a) 6.5 GeV and 8.2 GeV
 as well as
 with a WIMP mass of (b) 4.8 GeV and 5.8 GeV,
 respectively.
}
\label{fig:nu_dRdQ-B8-0000-05000-00000}
\end{figure}

 In Figs.~\ref{fig:nu_dRdQ-B8-0000-05000-00000},
 we show
 the dash--double--dotted magenta histograms
 with
 the nuclear recoil spectra of \CEnuNS\
 (the dashed red curves,
  shown in Figs.~\ref{fig:dRnuNSMdQ})
 off (a) $\rmF$
 and (b) $\rmI$ target nuclei,
 respectively,
 induced by Solar $\rmB$ neutrinos.
 One can find
 a clear discrepancy between
 the theoretical predictions
 and the realistic simulations:
 the event rate
 in the lowest energy bin
 would be significantly underestimated
 by around 70\%,
 whereas
 the event rates
 in the other (higher) energy bins
 would in contrast be overestimated
 with a factor of about 2.

 As a comparison,
 we draw
 two elastic WIMP--nucleus scattering spectra
 estimated theoretically
 by Eq.~\ref{eqn:dRdQ_SISD}
 with a relatively lighter WIMP mass
 (of 6.5 GeV for $\rmF$
  and of 4.8 GeV for $\rmI$,
  respectively)
 for fitting roughly
 the simulation histogram
 as well as
 with a relatively heavier WIMP mass
 (of 8.2 GeV for $\rmF$
  and of 5.8 GeV for $\rmI$,
  respectively)
 for fitting
 the theoretically predicted \CEnuNS\ spectrum.
 While
 the WIMP scattering spectra
 with the heavier masses
 can match the neutrino scattering spectra
 almost perfectly,
 except for
 the range around $Q = 0$,
 there is always some differences
 between the fitting WIMP scattering spectra
 with the lighter masses
 and the simulated \CEnuNS\ histograms,
 which can not be eliminated
 by simply modifying the WIMP masses.

\subsection[Angular distributions of
            the nuclear recoil flux/energy
            in the $\nuin$ coordinate system]
           {\boldmath
            Angular distributions of
            the nuclear recoil flux/energy
            in the $\nuin$ coordinate system}
\label{subsec:NR_ang-nu}

 In this subsection,
 we discuss first
 the angular distributions of
 the nuclear recoil flux
 and the accumulated and average recoil energies
 in the incoming--neutrino coordinate system.

 In the left column
 of Figs.~\ref{fig:nu_NR_theta-F19-B8-0000_0350-05000-\PlotNumberAa},
 we show
 the angular distributions of
 the nuclear recoil flux (top),
 the accumulated (middle)
 and the average (bottom) recoil energies of
 the light target nucleus $\rmF$
 in unit of the all--sky average values.
 (Note that
  the scale of the color bar
  used in the bottom frame is different
  from those in the top and the middle frames.)
 Meanwhile,
 the corresponding event number (top),
 the accumulated (middle)
 and the average (bottom) recoil energies of
 $\rmF$ target
 as functions of the equivalent recoil angle $\thetaNRnu$
 have also been provided
 in the right column,
 where
 the thin vertical dash--double--dotted magenta lines
 indicate the 1$\sigma$ statistical uncertainties.
 500 accepted \CEnuNS\ events on average
 in one entire year
 have been recorded
 and binned into 12 $\times$ 12 bins
 for the elevation and the azimuthal angle (left)
 and 12 bins in the range of $\thetaNRnu = 0$ and 90$^{\circ}$ (right).

 \def \Target       {F19}
 \def \ShortFrame   {in}
 \def \EnergyWindow {0000_0350}
 \def \PlotNumber   {\PlotNumberAa}
 \InsertFigureNRtheta
  {nu_}
  {\nuSource}
  {Left:
   the angular distributions of
   the nuclear recoil flux (top),
   the accumulated (middle)
   and the average (bottom) recoil energies of
   $\rmF$ target
   in unit of the all--sky average values.
   (Note that
    the scale of the color bar
    used in the bottom frame is different
    from those in the top and the middle frames.)
   Right:
   the dependence of
   the corresponding event number (top),
   the accumulated (middle)
   and the average (bottom) recoil energies of
   $\rmF$ target
   on the equivalent recoil angle $\thetaNRnu$;
   the thin vertical dash--double--dotted magenta lines
   indicate the 1$\sigma$ statistical uncertainties.
   500 accepted \CEnuNS\ events on average
   in one entire year
   have been recorded
   and binned into 12 $\times$ 12 bins
   for the elevation and the azimuthal angle (left)
   and 12 bins in the range of $\thetaNRnu = 0$ and 90$^{\circ}$ (right).
   See the text for further details.%
   \vspace{-0.4 cm}%
   }
 \def \Target       {I127}
 \def \EnergyWindow {0000_0050}
 \InsertFigureNRtheta
  {nu_}
  {\nuSource}
  {As in Figs.~\ref{fig:nu_NR_theta-F19-B8-0000_0350-05000-\PlotNumberAa},
   except that
   a heavy nucleus $\rmI$
   has been considered as our target.%
   }

 As a reference,
 in the plots
 shown in the right column,
 we draw also
 the dash--dotted blue and dashed purple curves
 to indicate two theoretical cases:
 all incident Solar $\rmB$ neutrinos
 have monotonically
 the root--mean--square energy
 $E_{\nu, {\rm rms}} = 7.26~{\rm MeV}$
 and
 the maximal--flux energy of 6.44 MeV,
 respectively.
 Since
 the nuclear recoil energy
 is approximately proportional to
 the neutrino incident energy squared,
 the former indicates
 the theoretical average of
 the transferable nuclear recoil energy
 as a function of the equivalent recoil angle.
 Additionally,
 the dotted green curve
 indicates that
 the approximation
 in the second line of
 the validation criterion (\ref{eqn:f_NR_thetaNRnu}) of
 $d\sigmanuNSM / d\thetaNRnu$
 has been used
 and the nuclear form factor suppressions
 have been ignored
 (i.e.,
  $\FSISDQ = 1$),
 instead of
 using the exact forms of
 $d\sigmanuNSM / dQ$,
 $dQ / d\eta$,
 and $Q (\Enu, \eta)$
 given by the second line of Eq.~(\ref{eqn:dsigma_nuN_dQQ})
 and the first lines of Eqs.~(\ref{eqn:dQQ_deta})
 and (\ref{eqn:QQ_eta})
 in our simulations.

 One can easily find that,
 as we calculated
 in Eqs.~(\ref{eqn:dsigma_nuN_dpsi:SI:eta:theta})
 as well as
 listed in Table \ref{tab:Q-dsigma_nuN_dpsi:eta:theta},
 the most frequent
 and the most energetic equivalent recoil angles
 are at $\thetaNRnu = 30^{\circ}$
 ($\eta = 60^{\circ}$
  \cite{OHare26})
 and    $\thetaNRnu = 45^{\circ}$,
 respectively.
 However,
 in the middle--right frame of
 Figs.~\ref{fig:nu_NR_theta-F19-B8-0000_0350-05000-\PlotNumberAa},
 one can find
 a tiny asymmetry
 with respect to $\thetaNRnu = 45^{\circ}$,
 which is caused by
 the much smaller but non--negligible contribution
 from the SD axial--vector interaction
 on the $\rmF$ nucleus
 in Eq.~(\ref{eqn:dsigma_nuN_dQQ}).
 Additionally,
 comparing to
 Figs.~\ref{fig:nu_NR_theta-I127-B8-0000_0050-05000-\PlotNumberAa},
 with a heavy nucleus $\rmI$
 as our target,
 the angular distributions of
 the nuclear recoil flux and energy
 shift in contrast a tiny bit
 towards smaller $\thetaNRnu$'s
 (larger $\eta$'s).
 This is caused by
 a stronger nuclear form factor suppression
 on high recoil energy
 (large $\thetaNRnu$
  or small $\eta$),
 even though
 the maximal transferable recoil energy
 to $\rmI$ (4.54 keV)
 is much smaller than that
 to $\rmF$ (30.27 keV)
 (see Fig.~\ref{fig:FQ}).

\begin{table} [t!]
\small
\begin{center}
\renewcommand{\arraystretch}{0.546}
\begin{tabular}{|| c | c | c | c ||}
\hline
\hline
 & & & \\
 \makebox[1.25 cm][c]{Angle}                           &
 \makebox[4.5  cm][c]{SI        vector term dominates} &
 \makebox[3.25 cm][c]{Two terms equal}                 &
 \makebox[5.72 cm][c]{SD axial--vector term dominates} \\
 & & & \\
\hline
\hline
 \multicolumn{4}{|| c ||}{} \\
 \multicolumn{4}{|| c ||}{Most frequent (equivalent) recoil angle} \\
 \multicolumn{4}{|| c ||}{} \\
\hline
 & & & \\
 $\eta$             &
 $60^{\circ}$       &
 $45^{\circ}$       &
 $   \frac{1}{2} \cos^{-1}\afrac{\sqrt{17} - 3}{4}
  =  36.85^{\circ}$ \\
 & & & \\
\hline
 & & & \\
 $\thetaNRnu$       &
 $30^{\circ}$       &
 $45^{\circ}$       &
 $   \frac{1}{2} \cos^{-1}\afrac{3 - \sqrt{17}}{4}
  =  53.15^{\circ}$ \\
 & & & \\
\hline
\hline
 \multicolumn{4}{|| c ||}{} \\
 \multicolumn{4}{|| c ||}{Most energetic (equivalent) recoil angle} \\
 \multicolumn{4}{|| c ||}{} \\
\hline
 & & & \\
 $\eta$             &
 $45^{\circ}$       &
 $30^{\circ}$       &
 $   \frac{1}{2} \cos^{-1}\afrac{\sqrt{73} - 5}{6}
  =  26.90^{\circ}$ \\
 & & & \\
\hline
 & & & \\
 $\thetaNRnu$       &
 $45^{\circ}$       &
 $60^{\circ}$       &
 $   \frac{1}{2} \cos^{-1}\afrac{5 - \sqrt{73}}{6}
  =  63.10^{\circ}$ \\
 & & & \\
\hline
\hline
\end{tabular}
\end{center}
\caption{
 The most frequent
 and the most energetic (equivalent) recoil angles
 $\eta$ ($\thetaNRnu$)
 in the incoming--neutrino coordinate system,
 estimated from
 Eqs.~(\ref{eqn:dsigma_nuN_deta:SI})
 and (\ref{eqn:Q-dsigma_nuN_deta:SI})
 as well as
 under the assumptions that
 the SD term
 in Eq.~(\ref{eqn:dsigma_nuN_dQQ})
 dominates
 and equals
 the SI term,
 respectively.
}
\label{tab:Q-dsigma_nuN_dpsi:eta:theta}
\end{table}

 Note here that,
 although
 the theoretical maximal transferable recoil energies of
 Solar $\rmB$ neutrinos
 to $\rmF$ and $\rmI$ nuclei
 are 30.27 keV and 4.54 keV,
 respectively,
 the simulated maximal average recoil energy per event
 in the largest equivalent recoil angle $\thetaNRnu$
 (the smallest recoil angle $\eta$)
 shown
 in the bottom--right frames of
 Figs.~\ref{fig:nu_NR_theta-F19-B8-0000_0350-05000-\PlotNumberAa}
 and   \ref{fig:nu_NR_theta-I127-B8-0000_0050-05000-\PlotNumberAa}
 are only 5.93 keV and \mbox{0.89 keV},
 respectively.
 These correspond to
 an monotonic neutrino incident energy of 7.23 MeV,
 which is just above
 the maximal--flux energy $E_{\nu, {\rm max. flux}} = 6.44~{\rm MeV}$
 and the average energy $\expv{\Enu} = 6.74~{\rm MeV}$ of
 Solar $\rmB$ neutrinos,
 and is equal to
 their root--mean--square energy
 $E_{\nu, {\rm rms}} = 7.26~{\rm MeV}$.
 This indicates that,
 firstly,
 the maximal average recoil energy per event
 would be
 the average of
 the estimated transferable recoil energy.
 Secondly and more importantly,
 the dash--dotted blue curves
 estimated with
 the 7.26--{\rm MeV} root--mean--square energy of
 Solar $\rmB$ neutrinos
 can describe
 the recoil angle and energy distributions
 shown in the right frames
 of Figs.~\ref{fig:nu_NR_theta-F19-B8-0000_0350-05000-\PlotNumberAa}
 and      \ref{fig:nu_NR_theta-I127-B8-0000_0050-05000-\PlotNumberAa}
 very well.

\subsubsection*{WIMP--nucleus scattering events}

 The recoil energy of
 the WIMP--scattered target nucleus
 and
 the differential WIMP--nucleus scattering cross section
 with respect to
 the equivalent recoil angle
 $\thetaNRchi$
 are given by
 \cite{DMDDD-3D-WIMP-N}
\beq
     Q
  =  \bbrac{\afrac{2 \mrN^2}{\mN} \vchiLab^2}
     \sin^2(\thetaNRchi)
\~,
\label{eqn:QQ_thetaNRchi}
\eeq
 and
\beq
     \Dd{\sigma_{\chi {\rm N}}}{\thetaNRchi}
  =  \bbigg{  \sigma_{\chi {\rm N}}^{\rm SI} \FSIQ
            + \sigma_{\chi {\rm N}}^{\rm SD} \FSDQ  }
     \sin(2 \thetaNRchi)
\~,
\label{eqn:dsigma_dthetaNRchi}
\eeq
 respectively.
 Here
 $\mrN$
 is the reduced mass of
 the WIMP mass $\mchi$ and
 that of the target nucleus $\mN$,
 $\vchiLab$ is
 the magnitude of
 the 3-D WIMP incident velocity
 in the laboratory coordinate system.
 $\sigma_{\chi {\rm N}}^{\rm (SI, SD)}$ are
 the SI/SD total cross sections
 ignoring the nuclear form factor suppression,
 respectively.
 It is then easy to solve that
 the most frequent
 and the most energetic equivalent recoil angles
 without the nuclear form factor suppression
 are at $\thetaNRchi = 45^{\circ}$
 and    $\thetaNRchi = 60^{\circ}$,
 respectively,
 both are larger than
 the $\thetaNRchi$'s in \CEnuNS.

\begin{table} [b!]
\begin{center}
\renewcommand{\arraystretch}{1.7}
\begin{tabular}{|| c || c | c | c  | c | c ||}
\hline
\hline
 \makebox[2.3 cm][c]{$\mchi$} &
 \makebox[2.3 cm][c]{$\rmF$}  &
 \makebox[2.3 cm][c]{$\rmAr$} &
 \makebox[2.3 cm][c]{$\rmGe$} &
 \makebox[2.3 cm][c]{$\rmXe$} &
 \makebox[2.3 cm][c]{$\rmW$}  \\
\hline
\hline
 \multicolumn{6}{|| c ||}{Most frequent equivalent recoil angle $\thetaNRchi$} \\
\hline
  20 GeV &
 $45.0^{\circ}$ &
 $44.5^{\circ}$ &
 $43.0^{\circ}$ &
 $40.5^{\circ}$ &
 $38.5^{\circ}$ \\
\hline
 100 GeV &
 $45.0^{\circ}$ &
 $42.0^{\circ}$ &
 $28.0^{\circ}$ &
 $22.0^{\circ}$ &
 $16.5^{\circ}$ \\
\hline
 200 GeV &
 $45.0^{\circ}$ &
 $37.0^{\circ}$ &
 $25.5^{\circ}$ &
 $15.5^{\circ}$ &
 $14.0^{\circ}$ \\
\hline
\hline
 \multicolumn{6}{|| c ||}{Most energetic equivalent recoil angle $\thetaNRchi$} \\
\hline
  20 GeV &
 $60.0^{\circ}$ &
 $59.0^{\circ}$ &
 $57.5^{\circ}$ &
 $56.0^{\circ}$ &
 $55.5^{\circ}$ \\
\hline
 100 GeV &
 $58.5^{\circ}$ &
 $52.5^{\circ}$ &
 $45.5^{\circ}$ &
 $35.0^{\circ}$ &
 $28.5^{\circ}$ \\
\hline
 200 GeV &
 $56.0^{\circ}$ &
 $49.0^{\circ}$ &
 $38.0^{\circ}$ &
 $27.0^{\circ}$ &
 $21.0^{\circ}$ \\
\hline
\hline
\end{tabular}
\end{center}
\caption{
 The most frequent
 and the most energetic equivalent recoil angles
 $\thetaNRchi$,
 estimated roughly from
 the simulation results
 presented in Ref.~\cite{DMDDD-NR}.
 Three different masses of
 incident Galactic halo WIMPs:
  20 GeV,
 100 GeV,
 and 200 GeV,
 as well as
 five target nuclei:
 $\rmF$,
 $\rmAr$,
 $\rmGe$,
 $\rmXe$,
 and $\rmW$,
 are considered.
}
\label{tab:Q-dsigma_dpsi:eta:theta}
\end{table}

 As summarized
 in Table \ref{tab:Q-dsigma_dpsi:eta:theta},
 the angular distributions of
 the nuclear recoil flux and energy
 induced by Galactic halo WIMPs
 (presented and discussed in detail
  in Ref.~\cite{DMDDD-NR})
 show high target--dependence:
 Due to
 the much stronger nuclear form factor suppression
 on high recoil energies
 and correspondingly large equivalent recoil angles $\thetaNRchi$
 (small recoil angles $\eta$),
 the heavier the mass of our target nucleus
 and/or that of incident Galactic halo WIMPs,
 the smaller (larger) the most frequent/energetic
 equivalent recoil angles $\thetaNRchi$
 (recoil angles $\eta$).

\subsection{Angular distributions of
            the nuclear recoil flux/energy
            in the laboratory coordinate system}
\label{subsec:NR_ang-Lab}

 In this subsection,
 we move to discuss
 the angular distributions of
 the nuclear recoil flux
 and the accumulated and average recoil energies
 induced by Solar $\rmB$ neutrinos
 and observed in the laboratory coordinate system%
\footnote{
 Similar works
 with non--directional direct DM/WIMP detectors
 can be found
 in e.g.~Ref.~\cite{YZhuang23}.
}.
 \def \Target       {F19}
 \def \ShortFrame   {Lab}
 \def \EnergyWindow {0000_0350}
 \InsertFigureNRangAnnual
  {nu_NR}
  {nu_Q}
  {nu_QoN}
  {\nuSource}
  {-\LabName}
  {The angular distributions of
   the nuclear recoil flux (top),
   the accumulated (middle)
   and the average (bottom) recoil energies of
   $\rmF$ target
   observed in the laboratory coordinate system of
   \LabName\
   \LabLocation\
   in unit of the all--sky average values.
   All simulation setup are
   as in Figs.~\ref{fig:nu_N_ang-B8-Lab-05000-\LabName}
   and         \ref{fig:nu_NR_theta-F19-B8-0000_0350-05000-\PlotNumberAa}.%
   }
 \def \Target       {I127}
 \def \EnergyWindow {0000_0050}
 \InsertFigureNRangAnnual
  {nu_NR}
  {nu_Q}
  {nu_QoN}
  {\nuSource}
  {-\LabName}
  {As in Figs.~\ref{fig:nu_NR_ang-F19-B8-Lab-0000_0350-05000-\LabName},
   except that
   a heavy nucleus $\rmI$
   has been considered as our target.%
   }

 In Figs.~\ref{fig:nu_NR_ang-F19-B8-Lab-0000_0350-05000-\LabName}
 and      \ref{fig:nu_NR_ang-I127-B8-Lab-0000_0050-05000-\LabName},
 we show
 the angular distributions of
 the nuclear recoil flux (top),
 the accumulated (middle)
 and the average (bottom) recoil energies of
 $\rmF$ and $\rmI$ targets
 observed at
 \LabName\
 \LabLocation\
 in unit of the all--sky average values,
 respectively.
 One entire year
 and four advanced 60-day seasons
 for the observation period of \CEnuNS\ events
 has been considered.

 First of all,
 due to
 the distributions of
 the recoil flux/energy
 peaked at the large (small equivalent) recoil angles
 (shown
  in the right columns of
  Figs.~\ref{fig:nu_NR_theta-F19-B8-0000_0350-05000-\PlotNumberAa}
  and   \ref{fig:nu_NR_theta-I127-B8-0000_0050-05000-\PlotNumberAa}),
 the angular distributions of
 the nuclear recoil flux
 and the accumulated recoil energy
 induced by Solar $\rmB$ neutrinos
 in the laboratory coordinate system
 would smear pretty widely.
 In contrast,
 the angular distribution of
 the average recoil energy
 concentrates clearly around the directions of incident neutrinos
 and thus show a clear annual variation
 (cf.~Figs.~\ref{fig:nu_N_ang-B8-Lab-05000-\LabName}).
 Secondly,
 as discussed in Sec.~\ref{subsec:NR_ang-nu},
 due to the low energy of
 incident Solar $\rmB$ neutrinos
 and thus the negligible nuclear form factor suppression,
 the angular distributions of
 the nuclear recoil flux and energy of
 different target nuclei
 would be experimentally indistinguishable.

\subsubsection*{WIMP--nucleus scattering events}

 In the Galactic point of view,
 the moving direction of halo WIMPs
 should be random and (approximately) isotropic.
 By combining with
 the movement of the Solar system
 around the Galaxy,
 the WIMP incident flux
 centers approximately at
 the direction
 from the Cygnus constellation
 to the Solar center
 (see Fig.~\ref{fig:v_Sun_Eq-S-G-rotated}).
 The orbital motion of the Earth around the Sun
 shift this main direction only negligibly.
 Hence,
 as presented and discussed
 in Ref.~\cite{DMDDD-NR},
 the angular recoil flux/energy distributions
 induced by Galactic halo WIMPs
 would almost be identical for
 one specified target nucleus
 at a fixed laboratory/location
 without an experimentally distinguishable annual variation,
 while
 the angular distributions of
 WIMP scattering events
 show a strong WIMP mass and target dependence.

\subsection{Angular distributions of
            the nuclear recoil flux/energy
            in the Equatorial coordinate system}
\label{subsec:NR_ang-Eq}

 We discuss
 in this subsection
 the angular distributions of
 the nuclear recoil flux
 and the accumulated and average recoil energies
 induced by Solar $\rmB$ neutrinos
 and observed in the Equatorial coordinate system.
 Note that,
 by definition,
 the Equatorial coordinate system
 is laboratory/location independent
 (see Sec.~\ref{appx:XYZ_G-S-Eq}).
 Hence,
 considering
 the very low theoretically estimated event rate,
 combining and analyzing \CEnuNS\ events
 off the same target nucleus
 from different underground laboratories
 in the Equatorial coordinate system
 would be a practically useful strategy.

 \def \Target       {F19}
 \def \ShortFrame   {Eq}
 \def \EnergyWindow {0000_0350}
 \InsertFigureNRangAnnual
  {nu_NR}
  {nu_Q}
  {nu_QoN}
  {\nuSource}
  {}
  {The angular distributions of
   the nuclear recoil flux (top),
   the accumulated (middle)
   and the average (bottom) recoil energies of
   $\rmF$ target
   observed in the Equatorial coordinate system
   in unit of the all--sky average values.
   All simulation setup are
   as in Figs.~\ref{fig:nu_N_ang-B8-Eq-05000}
   and         \ref{fig:nu_NR_ang-F19-B8-Lab-0000_0350-05000-\LabName}.%
   }
 \def \Target       {I127}
 \def \EnergyWindow {0000_0050}
 \InsertFigureNRangAnnual
  {nu_NR}
  {nu_Q}
  {nu_QoN}
  {\nuSource}
  {}
  {As in Figs.~\ref{fig:nu_NR_ang-F19-B8-Eq-0000_0350-05000},
   except that
   a heavy nucleus $\rmI$
   has been considered as our target.%
   }

 In Figs.~\ref{fig:nu_NR_ang-F19-B8-Eq-0000_0350-05000}
 and      \ref{fig:nu_NR_ang-I127-B8-Eq-0000_0050-05000},
 we show
 the angular distributions of
 the nuclear recoil flux (top),
 the accumulated (middle)
 and the average (bottom) recoil energies of
 $\rmF$ and $\rmI$ targets
 observed in the Equatorial coordinate system
 in unit of the all--sky average values,
 respectively.
 One entire year
 and four advanced 60-day seasons
 for the observation period of \CEnuNS\ events
 has been considered.
 It can clearly be found that
 the angular distributions of
 the nuclear recoil flux/energy
 in four advanced seasons
 vary annually
 along the trajectory of
 the moving direction of
 incident Solar $\rmB$ neutrinos.
 Corresponding to
 the distributions of
 the recoil flux/energy
 peaked at the large (small equivalent) recoil angles
 shown
 in the right columns of
 Figs.~\ref{fig:nu_NR_theta-F19-B8-0000_0350-05000-\PlotNumberAa}
 and   \ref{fig:nu_NR_theta-I127-B8-0000_0050-05000-\PlotNumberAa},
 one can observe here
 ring--like distributions of
 the most frequent
 and the most energetic recoil directions
 as well as
 concentrated distributions of
 the average recoil energy,
 respectively.

\subsubsection*{WIMP--nucleus scattering events}

 As mentioned above,
 the WIMP incident flux
 in the Equatorial and the Galactic coordinate systems
 centers approximately at
 the direction
 from the Cygnus constellation
 to the Solar center
 (see Fig.~\ref{fig:v_Sun_Eq-S-G-rotated}).
 The orbital motion of the Earth around the Sun
 shift this main direction only slightly
 and clockwise around the center at
 the direction
 from the Cygnus constellation
 \cite{DMDDD-N}.
 Hence,
 as presented and discussed
 in Ref.~\cite{DMDDD-NR},
 the distributions of
 the WIMP--scattered nuclear recoil flux and energy
 are basically fixed
 around the direction opposite to
 the Solar Galactic movement
 in the Equatorial coordinate system
 \cite{Bandyopadhyay10}:
 42.00$^{\circ}$S, 50.70$^{\circ}$W.
 Remind that
 the patterns of
 the WIMP--induced recoil flux/energy distributions
 are WIMP--mass and target dependent.

\subsection{Angular distributions of
            the nuclear recoil flux/energy
            in the geocentric Galactic coordinate system}
\label{subsec:NR_ang-geoG}

 For the sake of completeness
 as well as
 for comparing with works
 presented in e.g.~Ref.~\cite{OHare15b},
 we present
 in this subsection
 the angular distributions of
 the nuclear recoil flux
 and the accumulated and average recoil energies
 induced by Solar $\rmB$ neutrinos
 and observed in the geocentric Galactic coordinate system.
 Since
 the geocentric Galactic coordinate system
 is by definition laboratory/location independent
 (see Sec.~\ref{appx:XYZ_G-S-Eq}),
 one can practically
 combine and analyze \CEnuNS\ events
 off the same target nucleus
 from different underground laboratories
 in the geocentric Galactic coordinate system.

 \def \Target       {F19}
 \def \ShortFrame   {G}
 \def \EnergyWindow {0000_0350}
 \InsertFigureNRangAnnual
  {nu_NR}
  {nu_Q}
  {nu_QoN}
  {\nuSource}
  {}
  {The angular distributions of
   the nuclear recoil flux (top),
   the accumulated (middle)
   and the average (bottom) recoil energies of
   $\rmF$ target
   observed in the geocentric Galactic coordinate system
   in unit of the all--sky average values.
   All simulation setup are
   as in Figs.~\ref{fig:nu_N_ang-B8-G-05000}
   and         \ref{fig:nu_NR_ang-F19-B8-Lab-0000_0350-05000-\LabName}.%
   }
 \def \Target       {I127}
 \def \EnergyWindow {0000_0050}
 \InsertFigureNRangAnnual
  {nu_NR}
  {nu_Q}
  {nu_QoN}
  {\nuSource}
  {}
  {As in Figs.~\ref{fig:nu_NR_ang-F19-B8-G-0000_0350-05000},
   except that
   a heavy nucleus $\rmI$
   has been considered as our target.%
   }

 In Figs.~\ref{fig:nu_NR_ang-F19-B8-G-0000_0350-05000}
 and      \ref{fig:nu_NR_ang-I127-B8-G-0000_0050-05000},
 we show
 the angular distributions of
 the nuclear recoil flux (top),
 the accumulated (middle)
 and the average (bottom) recoil energies of
 $\rmF$ and $\rmI$ targets
 observed in the geocentric Galactic coordinate system
 in unit of the all--sky average values,
 respectively.
 One entire year
 and four advanced 60-day seasons
 for the observation period of \CEnuNS\ events
 has been considered.
 The same as those
 in the Equatorial coordinate system,
 the angular distributions of
 the nuclear recoil flux/energy
 in four advanced seasons
 vary annually
 along the trajectory of
 the moving direction of
 incident Solar $\rmB$ neutrinos.
 Distorted ring--like distributions of
 the most frequent
 and the most energetic recoil directions
 as well as
 the concentrated distributions of
 the average recoil energy
 can also be observed.

\section{Summary}

 In this paper,
 we investigated
 3-dimensional coherent elastic neutrino--nucleus scattering
 and simulated
 3-D nuclear recoil response of
 directional direct Dark Matter detectors
 induced by Solar $\rmB$ neutrinos.

 Firstly,
 we confirmed that,
 in the Standard Model,
 the most frequent
 and the most energetic recoil angles
 are at
 the recoil angles of $\eta = 60^{\circ}$
 and $\eta = 45^{\circ}$,
 respectively.
 Correspondingly,
 in the Equatorial
 and the geocentric Galactic coordinate systems,
 one could experimentally observe
 the characteristic ring--like distributions of
 the most frequent
 and the most energetic recoil directions
 as well as
 the concentrated distributions of
 the average recoil energy.

 Moreover,
 in the laboratory
 as well as
 in the Equatorial
 and the geocentric Galactic coordinate systems,
 the angular distributions of
 the nuclear recoil flux
 as well as
 the accumulated and the average recoil energies of
 \CEnuNS\ events
 show clearly the annual variations (shifts)
 around/along the moving direction of
 incident Solar neutrinos,
 while
 the patterns of
 the recoil flux/energy distributions
 induced by Galactic halo WIMPs
 would almost be fixed
 around
 the direction opposite to
 the Solar Galactic movement
 with in contrast clear
 WIMP--mass and target dependence.
 However,
 the peaks of
 the most frequent
 and the most energetic recoil angles
 as well as
 the patterns of
 the ring--like and the concentrated
 angular recoil flux/energy distributions
 depend on the relative strength of
 the spin--independent vector
 and the spin--dependent axial--vector neutrino--nucleus interactions.
 Hence,
 (precise) experimental observations of
 the direction--sensitive DM/WIMP--neutrino detectors
 in the future
 could provide information on
 (non--)standard neutrino--nucleus interactions.

 It is necessary to remind here that,
 firstly,
 due to the low energy of
 incident Solar $\rmB$ neutrinos
 and thus the negligible nuclear form factor suppression,
 in the near future
 with only a few tens or even hundreds of \CEnuNS\ events,
 the patterns of
 the angular recoil flux/energy distributions of
 different target nuclei
 would be experimentally indistinguishable.
 Secondly,
 the maximal average recoil energy per event
 induced by Solar $\rmB$ neutrinos
 with a maximal incident energy of \mbox{16.36 MeV}
 could be as low as only
 5.93 keV (for $\rmF$) or even 0.89 keV (for $\rmI$),
 which are only $\sim$ 20\% of
 the theoretically estimated
 maximal transferable recoil energies of
 Solar $\rmB$ neutrinos
 to $\rmF$ (30.27 keV) or $\rmI$ (4.54 keV) nuclei.

 In our simulations
 presented in this paper,
 500 accepted \CEnuNS\ events
 on average
 in one observation period
 (365 days/year
  or 60 days/season)
 in one experiment
 for one laboratory/target nucleus
 have been considered.
 In at least this and probably the next decades,
 even $\cal O$(500) total \CEnuNS\ signals
 (per year) to observe
 (in one experiment/underground laboratory)
 would be a strong challenge
 for all (directional) direct Dark Matter detectors.
 As a practical strategy,
 one could combine
 the recorded 3-D recoil information of \CEnuNS\ events
 offered by different underground laboratories
 with the same detector material
 in the laboratory/location--independent
 Galactic and Equatorial coordinate systems
 for analyses.

 Regarding the observation periods
 considered in our simulations,
 we used several approximations
 about the Earth's orbital motion
 in the Solar system.
 First,
 the Earth's orbit around the Sun is perfectly circular
 in the Ecliptic plane
 and the orbital speed is thus a constant.
 Second,
 the date of the vernal equinox is exactly fixed
 at the end of the March 20th (the 79th day) of
 a 365-day year
 and the few extra hours
 in an actual Solar year
 have been neglected.
 Nevertheless,
 considering the low \CEnuNS\ event rate
 and thus maximal a few hundreds of total (combined) \CEnuNS\ events
 observed in at least a few tens (or even hundreds) of days
 for the first--phase analyses,
 these approximations should be acceptable.

 In summary,
 we have
 extended our
 double--Monte Carlo scattering--by--scattering simulation
 for WIMP--nucleus scattering
 to neutrino--nucleus scattering.
 Hopefully,
 this works
 could have direct implications
 for the design and interpretations of
 future directional direct Dark Matter/WIMP and/or neutrino detection experiments
 as well as
 provide our colleagues
 a useful approach
 for studying neutrino physics
 beyond the Standard Model.

\subsubsection*{Acknowledgments}

 CLS appreciates
 N.~Bozorgnia and P.~Gondolo
 for useful discussions
 about the transformations between the celestial coordinate systems.
 CLS would also like to thank
 the friendly hospitality of
 the Institute for Theoretical Physics,
 School of Physics and Optoelectronic Engineering,
 Beijing University of Technology
 and
 the School of Nuclear Science and Technology,
 Lanzhou University
 as well as
 the pleasant atmosphere of
 the W101 Ward and the Cancer Center of
 the Kaohsiung Veterans General Hospital,
 where part of this work was completed.

\appendix
\setcounter{equation}{0}
\setcounter{figure}  {0}
\setcounter{table}   {0}
 \renewcommand{\theequation}{A\arabic{equation}}
 \renewcommand{\thefigure}  {A\arabic{figure}}
 \renewcommand{\thetable}   {A\arabic{table}}
%

%
%
%
 \renewcommand{\arraystretch}{1.5}
\section{Definitions of and transformations between our coordinate systems}
\label{appx:XYZ}

 In this section,
 we review briefly
 our definitions of
 the laboratory--independent
 (Galactic,
  Ecliptic,
  Equatorial,
  and Earth)
 coordinate systems
 as well as
 the laboratory--dependent
 (horizontal
  and laboratory)
 coordinate systems
 used in our simulation package.
 We also summarize
 the transformations of
 the generated 3-D moving direction of
 Solar $\rmB$ neutrinos
 and
 the recoil direction of
 the neutrino--scattered target
 between different celestial coordinate systems
 as well as
 the matrices
 needed for the transformations
 between these coordinate systems.

 Discussions about our coordinate systems
 as well as
 the detailed derivations of the transformation matrices
 between them
 can be found in Ref.~\cite{DMDDD-N}.

\subsection{Laboratory--independent coordinate systems}
\label{appx:XYZ_G-S-Eq}

 We consider first
 the Galactic,
 the Ecliptic,
 and the Equatorial coordinate systems.

\subsubsection{Definitions}
\label{appx:def-XYZ_G-S-Eq}

 In Fig.~\ref{fig:v_Sun_Eq-S-G-rotated},
 we show
 the definitions of
 and the relative orientations between
 the (black) Galactic,
 the (red) Ecliptic,
 and the (blue) Equatorial coordinate systems
 (on the date of the vernal equinox).

\InsertSKPPlotS
 {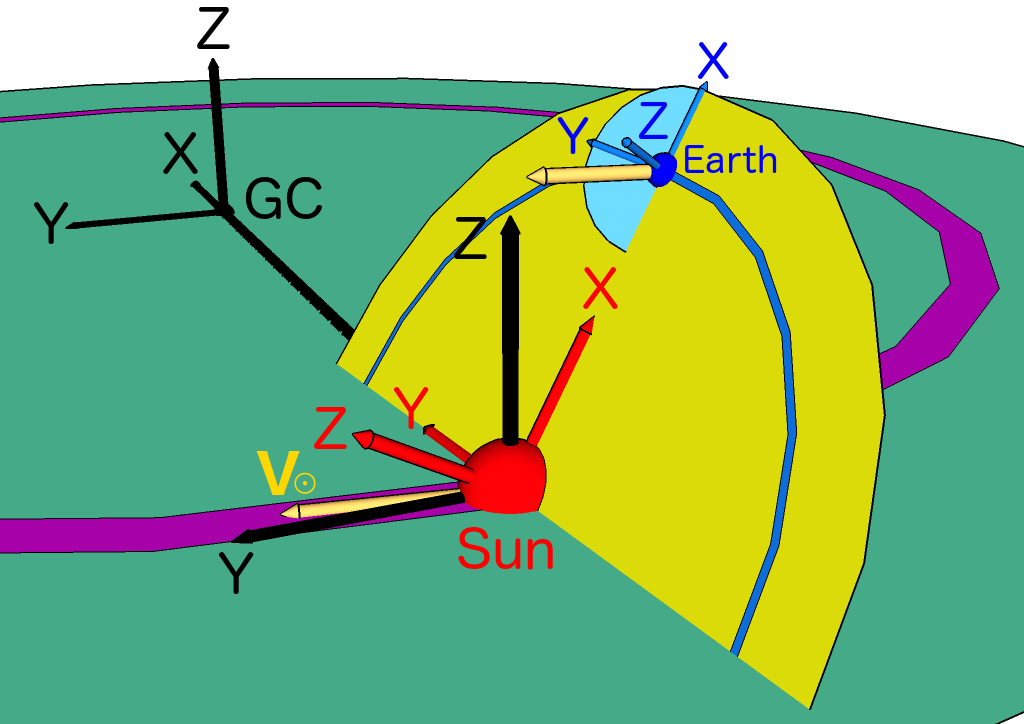}
 {The definitions of
  and the relative orientations between
  the (black) Galactic,
  the (red) Ecliptic,
  and the (blue) Equatorial coordinate systems
  (on the date of the vernal equinox).
  While
  the magenta circular band
  indicates an approximate path of
  the orbital motion of the Solar system
  in the Galaxy
  and
  the blue circular band
  the Earth's orbit around the Sun,
  the additional (golden) arrows
  indicate
  the direction of the movement of the Solar system
  around the Galactic center,
  which towards the Cygnus constellation.
  See the text for the detailed descriptions.%
  }

 Firstly,
 the origin of the Galactic coordinate system
 is at the (approximate) Galactic Center (GC).
 The primary direction (the $\xG$--axis)
 points from the Solar center to GC
 and
 the $\zG$--axis
 to the Galactic North Pole (GNP).
 Then
 the right--handed convention
 is used for defining the $\yG$--axis
 and the fundamental ($\xG - \yG$) plane
 is the approximate Galactic plane.

 Meanwhile,
 the origins of
 the Ecliptic
 and the Equatorial coordinate systems
 are at
 the center of the Sun
 and that of the Earth,
 respectively.
 The common primary direction (the $\xS$/$\xEq$--axis)
 is the direction
 pointing from the Solar center to the Earth's center
 at 12 o'clock midnight (the end) of
 the date of the vernal equinox,
 which
 is fixed exactly
 at the 79th day (the March 20th) of
 a 365-day year,
 the $\zS$-- and the $\zEq$--axes
 are perpendicular to the (yellow) Ecliptic
 and the (blue) Equatorial planes,
 respectively,
 and their $\yS$-- and $\yEq$--axes
 are then defined as usual by
 the right--handed convention.

 Additionally,
 in Fig.~\ref{fig:v_Sun_Eq-S-G-rotated},
 we also draw
 two (golden) arrows
 to indicate
 the direction of the movement of the Solar system
 around the Galactic center,
 which towards the Cygnus constellation.
 Note that
 the moving direction of the Solar system
 is not parallel to,
 but only approximately along
 the $\yG$--axis,
 (with an included angle of 8.87$^{\circ}$),
 nor in the (approximate) Galactic plane
 (0.60$^{\circ}$ above).

 Note also that
 the Ecliptic coordinate system
 only moves approximately linearly
 with the Solar Galactic orbital speed
 $\vSunG = |\bfvSunG| \simeq 220$ km/s
 and
 the tiny Galactic orbital rotation of
 the Solar system
 is considered to be imperceptible,
 whereas
 the Equatorial coordinate system
 moves orbitally around
 (and also linearly with) the Sun,
 but doesn't rotate.
 These mean that
 the axes of
 the Galactic,
 the Ecliptic,
 and the Equatorial coordinate systems
 defined in our simulation package
 are all fixed
 (see Table \ref{tab:CSs}
  for the summary of
  the styles of
  the movements and the rotations of
  different celestial coordinate systems).

\subsubsection{Transformations
               between the geocentric Galactic
               and the geocentric Ecliptic coordinate systems}
\label{appx:tr-G-S}

 In our simulation package,
 the 3-D momentum of
 Solar $\rmB$ neutrinos
 generated in the Ecliptic coordinate system,
 ${\bf p}_{\nu, {\rm S}}$,
 as well as
 the recoil direction of
 the $\nu$--scattered target nucleus
 can be transformed
 into the geocentric Galactic coordinate system%
\footnote{
 The coordinate axes of
 the geocentric Galactic and Ecliptic coordinate systems
 are parallel to
 our Galactic and Ecliptic coordinate systems,
 respectively,
 but the origins are located
 at the Earth's center.
 Hence,
 the transformations
 between our Equatorial and geocentric Galactic/Ecliptic coordinate systems
 are pure rotations
 and the Earth's velocities
 in the Galaxy and in the Solar system
 are discarded.
}
 by
\cheqnXa{A}
\beq
     {\bf p}_{\nu, {\rm G}}
  =  \MaSG \~ {\bf p}_{\nu, {\rm S}}
\~,
\label{eqn:bfpnuS->bfpnuG}
\eeq
 and
\cheqnXb{A}
\beq
     {\bf v}_{\rm N_R, G}
  =  \MaSG \~ {\bf v}_{\rm N_R, S}
\~,
\label{eqn:bfvNRS->bfvNRG}
\eeq
\cheqnX{A}
 respectively,
 with
 the transformation matrix $\MaSG$
 given in Eq.~(\ref{eqn:Ma_S_G}).

\subsubsection{Transformations
               between the geocentric Ecliptic
               and the Equatorial coordinate systems}
\label{appx:tr-S-Eq}

 Similar to Eq.~(\ref{eqn:bfpnuS->bfpnuG}),
 the 3-D momentum of
 Solar $\rmB$ neutrinos
 in the Equatorial coordinate system
 can be obtained by
 transforming ${\bf p}_{\nu, {\rm S}}$
 in the Ecliptic coordinate system:
\cheqnXa{A}
\beq
     {\bf p}_{\nu, {\rm Eq}}
  =  \MaSEq \~ {\bf p}_{\nu, {\rm S}}
\~,
\label{eqn:bfpnuS->bfpnuEq}
\eeq
 with
 the transformation matrix $\MaSEq$
 given in Eq.~(\ref{eqn:Ma_S_Eq}).
 Conversely,
 for transforming
 the recoil direction of
 the $\nu$--scattered target nucleus
 into the geocentric Ecliptic coordinate system,
 we have
\cheqnXb{A}
\beq
     {\bf v}_{\rm N_R, S}
  =  \MaEqS \~ {\bf v}_{\rm N_R, Eq}
\~,
\label{eqn:bfvNREq->bfvNRS}
\eeq
\cheqnX{A}
 with the transformation matrix $\MaEqS$
 given in Eq.~(\ref{eqn:Ma_Eq_S}).

\subsubsection{Transformation matrices}
\label{appx:Ma-XYZ_G-S-Eq}

 Firstly,
 by definition,
 the transformation matrix
 from the Ecliptic coordinate system
 to the Equatorial coordinate system
 can be given directly as
\cheqnXa{A}
\beqn
     \MaSEq
  =  \left[\begin{array}{c c c}
             1 ~&~   0              ~&~  0              \\
             0 ~&~  \cos(\psiEarth) ~&~ \sin(\psiEarth) \\
             0 ~&~ -\sin(\psiEarth) ~&~ \cos(\psiEarth) \\
           \end{array}\right]
  =  \left[\begin{array}{c c c}
             1 ~&~  0       ~&~ 0       \\
             0 ~&~  0.91775 ~&~ 0.39715 \\
             0 ~&~ -0.39715 ~&~ 0.91775 \\
           \end{array}\right]
\~,
\label{eqn:Ma_S_Eq}
\eeqn
 and,
 conversely,
\cheqnXb{A}
\beq
     \MaEqS
  =  \left[\begin{array}{c c c}
             1 ~&~  0              ~&~   0              \\
             0 ~&~ \cos(\psiEarth) ~&~ -\sin(\psiEarth) \\
             0 ~&~ \sin(\psiEarth) ~&~  \cos(\psiEarth) \\
           \end{array}\right]
  =  \left[\begin{array}{c c c}
             1 ~&~ 0       ~&~  0       \\
             0 ~&~ 0.91775 ~&~ -0.39715 \\
             0 ~&~ 0.39715 ~&~  0.91775 \\
           \end{array}\right]
\~,
\label{eqn:Ma_Eq_S}
\eeq
\cheqnX{A}
 where
 $\psiEarth = 23.4^{\circ}$
 is the Earth's obliquity.

 On the other hand,
 the directions of the Galactic Center
 and the Galactic North Pole
 in the Equatorial coordinate system
 can be expressed by
\cheqnXa{A}
\beqn
     \xGEq
 \=  \left[\begin{array}{c c c}
            \cos(\thetaGC) \cos(\phiGC) ~&~
            \cos(\thetaGC) \sin(\phiGC) ~&~
            \sin(\thetaGC)               \\
           \end{array}\right]_{\rm Eq}
     \non\\
 \eqncong
     \left[\begin{array}{c c c}
             0.05495 ~&~
             0.87340 ~&~
            -0.48389  \\
           \end{array}\right]_{\rm Eq}
\~,
\label{eqn:GC_Eq}
\eeqn
 and
\cheqnXb{A}
\beqn
     \zGEq
 \=  \left[\begin{array}{c c c}
            \cos(\thetaGNP) \cos(\phiGNP) ~&~
            \cos(\thetaGNP) \sin(\phiGNP) ~&~
            \sin(\thetaGNP)                \\
           \end{array}\right]_{\rm Eq}
     \non\\
 \eqncong
      \left[\begin{array}{c c c}
            0.86769 ~&~
            0.19793 ~&~
            0.45601  \\
           \end{array}\right]_{\rm Eq}
\~,
\label{eqn:GNP_Eq}
\eeqn
\cheqnX{A}%
 respectively,
 where
 we have adopted the values
 provided by Ref.~\cite{Wiki-Galactic}%
\footnote{
 Note that
 the common $\xS$/$\xEq$--axis
 defined in our Ecliptic and Equatorial coordinate systems
 points from the center of the Sun to that of the Earth
 and is thus opposite to
 the conventional astronomical definition.
 Hence,
 the right ascensions of GC
 and GNP
 in the Equatorial coordinate system
 given here
 differ from the values given in Ref.~\cite{Wiki-Galactic}
 by $180^{\circ}$.
\label{fn:phi_GC_Eq}
}
\cheqnXa{A}
\beq
     \thetaGC
  = -28.94^{\circ}
\~,
     ~~~~ ~~~~ ~~~~ ~~~~ ~~~~ 
     \phiGC
  =  86.40^{\circ}
\~,
\label{eqn:ang_GC_Eq}
\eeq
 and
\cheqnXb{A}
\beq
     \thetaGNP
  =  27.13^{\circ}
\~,
     ~~~~ ~~~~ ~~~~ ~~~~ ~~~~ ~ 
     \phiGNP
  =  12.85^{\circ}
\~,
\label{eqn:ang_GNP_Eq}
\eeq
\cheqnX{A}
 as the declinations and the right ascensions of
 GC
 and GNP
 in the Equatorial coordinate system,
 respectively.
 Then,
 by combining Eqs.~(\ref{eqn:GC_Eq}) and (\ref{eqn:GNP_Eq}),
 the $\yG$--axis of the Galactic coordinate system
 in the Equatorial coordinate system
 can be calculated by
\cheqnXNx{A}{-2}{c}
\beqn
     \yGEq
 \=  \zGEq \times \xGEq
     \non\\
 \=  \footnotesize
     \left[\renewcommand{\arraystretch}{1.75}
           \begin{array}{c}
              \cos(\thetaGNP) \sin(\phiGNP)
              \sin(\thetaGC)
            - \sin(\thetaGNP)
              \cos(\thetaGC)  \sin(\phiGC)  \\
              \sin(\thetaGNP)
              \cos(\thetaGC)  \cos(\phiGC)
            - \cos(\thetaGNP) \cos(\phiGNP)
              \sin(\thetaGC)                \\
              \cos(\thetaGNP) \cos(\thetaGC)
              \bBig{\cos(\phiGNP) \sin(\phiGC) - \sin(\phiGNP) \cos(\phiGC)} \\
           \end{array}\right]_{\rm Eq}^{\rm T}
     \non\\
 \eqncong
      \left[\begin{array}{c c c}
            -0.49406 ~&~
             0.44492 ~&~
             0.74696  \\
           \end{array}\right]_{\rm Eq}
\~.
\label{eqn:Y_G_Eq}
\eeqn
\cheqnXN{A}{1}%
 Hence,
 the transformation matrices
 between the Equatorial
 and the Galactic coordinate systems
 can be given by
\cheqnXa{A}
\beq
     \MaEqG
  =  \left[\begin{array}{c}
            \xGEq \\
            \yGEq \\
            \zGEq \\
           \end{array}\right]
  =  \left[\begin{array}{c c c}
             0.05495 ~&~ 0.87340 ~&~ -0.48389 \\
            -0.49406 ~&~ 0.44492 ~&~  0.74696 \\
             0.86769 ~&~ 0.19793 ~&~  0.45601 \\
           \end{array}\right]
\~,
\label{eqn:Ma_Eq_G}
\eeq
 and,
 conversely,
\cheqnXb{A}
\beq
     \MaGEq
  =  \left[\begin{array}{c c c}
             0.05495 ~&~ -0.49406 ~&~ 0.86769 \\
             0.87340 ~&~  0.44492 ~&~ 0.19793 \\
            -0.48389 ~&~  0.74696 ~&~ 0.45601 \\
           \end{array}\right]
\~.
\label{eqn:Ma_G_Eq}
\cheqnX{A}
\eeq

 Finally,
 by combining with
 the transformation matrices
 between the Equatorial
 and the Ecliptic coordinate systems
 in Eqs.~(\ref{eqn:Ma_Eq_S}) and (\ref{eqn:Ma_S_Eq}),
 the transformations
 between the Galactic
 and the Ecliptic coordinate systems
 can be obtained by
\cheqnXa{A}
\beq
     \MaGS
  =  \MaEqS \~ \MaGEq
  =  \left[\begin{array}{c c c}
             0.05495 ~&~ -0.49406 ~&~ 0.86769 \\
             0.99374 ~&~  0.11168 ~&~ 0.00055 \\
            -0.09723 ~&~  0.86223 ~&~ 0.49711 \\
           \end{array}\right]
\~,
\label{eqn:Ma_G_S}
\eeq
 and,
 conversely,
\cheqnXb{A}
\beq
     \MaSG
  =  \MaEqG \~ \MaSEq
  =  \left[\begin{array}{c c c}
             0.05495 ~&~ 0.99374 ~&~ -0.09723 \\
            -0.49406 ~&~ 0.11168 ~&~  0.86223 \\
             0.86769 ~&~ 0.00055 ~&~ 0.49711  \\
           \end{array}\right]
\~.
\label{eqn:Ma_S_G}
\eeq
\cheqnX{A}
\subsection{Earth coordinate system}
\label{appx:XYZ_E}

 As the connection of
 the Equatorial and the Ecliptic coordinate systems
 to the horizontal and the laboratory coordinate systems
 \cite{Bandyopadhyay10},
 we defined the Earth coordinate system
 in our simulation package.

\subsubsection{Definition}
\label{appx:def-XYZ_E}
\InsertSKPPlotS
 {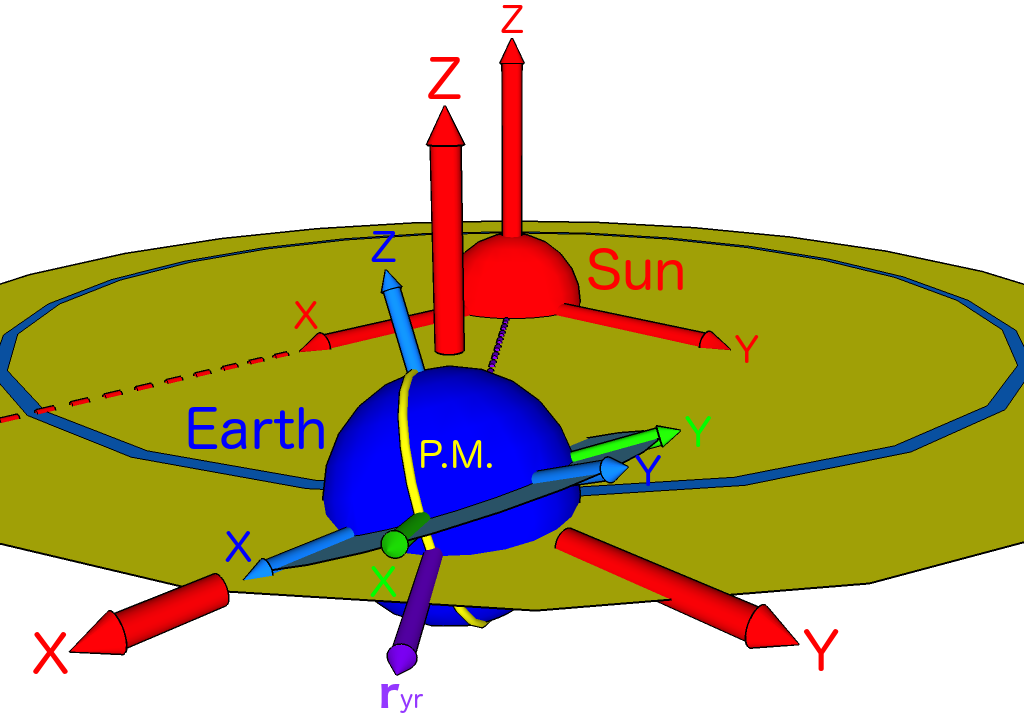}
 {The definition of
  the (light--green) Earth coordinate system
  at 12 o'clock midnight (the beginning)
  (i.e.,
   when the (yellow) Prime Meridian
   (the longitude 0$^{\circ}$)
   passes the purple arrow $\ryear$
   pointing from the Solar center to the Earth's center)
  of each single Solar day.
  The (red) Ecliptic and the (blue) Equatorial coordinate systems
  as well as
  the (blue) Earth's orbit around the Sun
  are also given here.
  See the text for the detailed description.%
  }

 As shown in Fig.~\ref{fig:E-Eq-S},
 we define
 the Earth coordinate system
 as follows:
 while
 the origin is also
 located at the Earth's center
 and
 the $\zE$--axis is still
 the Earth's north polar axis,
 the primary direction (the $\xE$--axis)
 points now from the Earth's center
 to the Prime Meridian
 (the longitude 0$^{\circ}$)
 at 12 o'clock midnight (the beginning)
 (i.e.,
  when the Prime Meridian
  passes the direction
  pointing from the Solar center to the Earth's center)
 of each single Solar day.
 The fundamental ($\xE - \yE$) plane is again
 the Equatorial plane
 and the right--hand convention is used
 to define the $\yE$--axis.

 Note that,
 for each single (Solar) day,
 the Earth coordinate system
 is fixed with the direction of the Prime Meridian
 at (UTC) 12 o'clock midnight,
 but rotates with the Earth
 during its orbital motion around the Sun.
 This means that
 our Earth coordinate system
 changes daily and discretely
 (see Table \ref{tab:CSs}).

\subsubsection{Transformations
               between the Equatorial and the Earth coordinate systems}
\label{appx:tr-Eq-E}

 In our simulation package,
 the transformations of
 the 3-D neutrino momentum
 at the UTC incoming/scattering time $t$
 between the Equatorial
 and the Earth coordinate systems
 are pure rotations,
 which can be given by
\cheqnXa{A}
\beq
     {\bf p}_{\nu, {\rm E}}
  =  \MaEqE (t) \~ {\bf p}_{\nu, {\rm Eq}}
\~,
\label{eqn:bfpnuEq->bfpnuE}
\eeq
 and,
 conversely,
 one has
\cheqnXb{A}
\beq
     {\bf v}_{\rm N_R, Eq}
  =  \MaEEq (t) \~ {\bf v}_{\rm N_R, E}
\~,
\label{eqn:bfvNRE->bfvNREq}
\eeq
\cheqnX{A}
 with the time--dependent transformation matrices
 $\MaEqE (t)$ and $\MaEEq (t)$
 given in Eqs.~(\ref{eqn:Ma_Eq_E}) and (\ref{eqn:Ma_E_Eq}),
 respectively.

\subsubsection{Transformation matrices}
\label{appx:Ma-XYZ_Eq-E}

 Following the calculations
 done by A.~Bandyopadhyay and D.~Majumdar
 in Ref.~\cite{Bandyopadhyay10},
 the transformation matrices
 between the Equatorial
 and the Earth coordinate systems
 can be expressed by
 \cite{DMDDD-N}
\cheqnXa{A}
\beqn
     \MaEqE (t)
 \=  \left[\begin{array}{c c c}
             \gamma (t) \cos(\psiyear (t - \tPM))                 ~&~
             \gamma (t) \sin(\psiyear (t - \tPM)) \cos(\psiEarth) ~&~
              0                                                    \\
            -\gamma (t) \sin(\psiyear (t - \tPM)) \cos(\psiEarth) ~&~
             \gamma (t) \cos(\psiyear (t - \tPM))                 ~&~
              0                                                    \\
              0                                                   ~&~
              0                                                   ~&~
              1                                                    \\
           \end{array}\right]
\~,
     \non\\
\label{eqn:Ma_Eq_E}
\eeqn
 and,
 conversely,
\cheqnXb{A}
\beqn
     \MaEEq (t)
 \=  \left[\begin{array}{c c c}
             \gamma (t) \cos(\psiyear (t - \tPM))                 ~&~
            -\gamma (t) \sin(\psiyear (t - \tPM)) \cos(\psiEarth) ~&~
              0                                                    \\
             \gamma (t) \sin(\psiyear (t - \tPM)) \cos(\psiEarth) ~&~
             \gamma (t) \cos(\psiyear (t - \tPM))                 ~&~
              0                                                    \\
              0                                                   ~&~
              0                                                   ~&~
              1                                                    \\
           \end{array}\right]
\~.
     \non\\
\label{eqn:Ma_E_Eq}
\eeqn
\cheqnX{A}%
 Here
 we define
\beq
         \gamma (t)
 \equiv  \frac{1}
              {\sqrt{  \cos^2(\psiyear (t - \tPM))
                     + \sin^2(\psiyear (t - \tPM)) \cos^2(\psiEarth) } }
\~,
\label{eqn:psi_yr_Earth}
\eeq
 the angle
 swept by the connection
 between the Solar and the Earth's centers
 from the day of the vernal equinox
 (the 79th day)
 can be expresses by
\beq
         \psiyear (t)
 \equiv  \frac{2 \pi}{365} \abrac{t - 79.0}
\~,
\label{eqn:psi_yr}
\eeq
 and
 $\tPM$ indicates
 the fractional part of
 the UTC incoming/scattering time $t$ of
 each recorded \CEnuNS\ event
 in unit of day
 (see Fig.~\ref{fig:E-H-Lab}(b)).

\subsection{Laboratory--dependent coordinate systems}
\label{appx:XYZ_H-Lab}

 Now
 we come to
 the horizontal and the laboratory coordinate systems.

\subsubsection{Definitions}
\label{appx:def-XYZ_H-Lab}
\InsertSKPPlotD
 {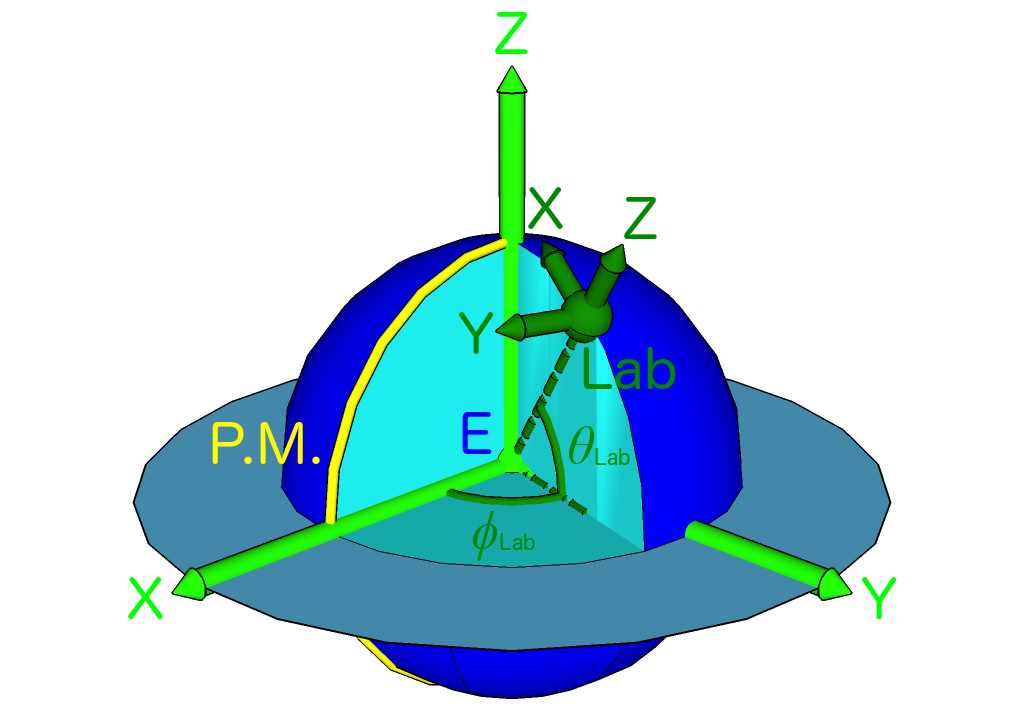}
 {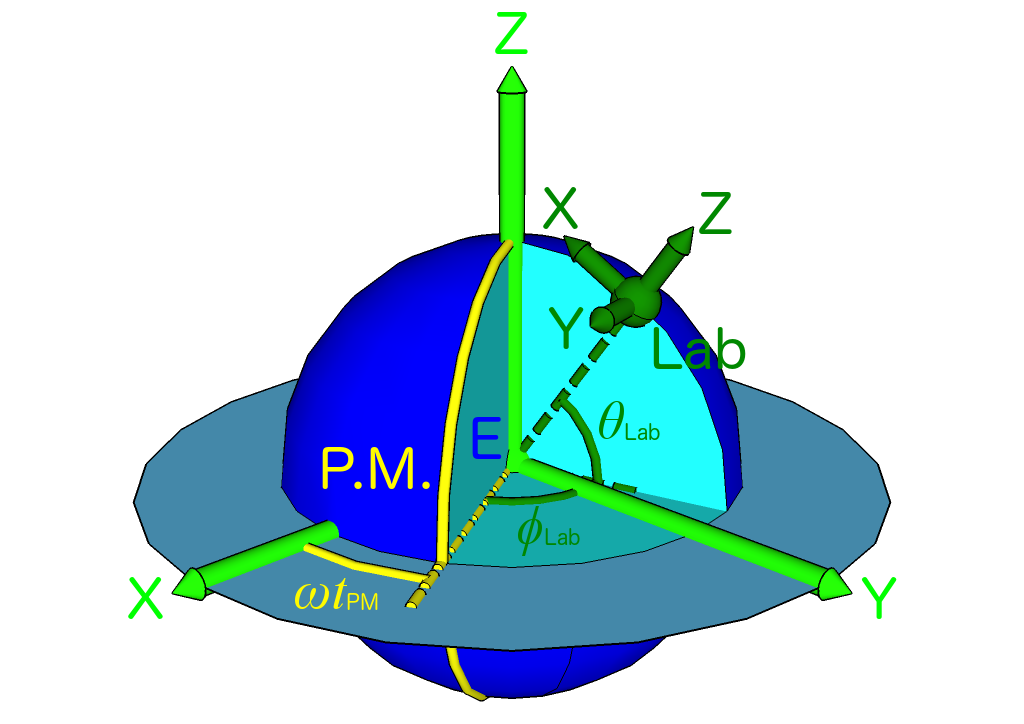}
 {Horizontal coordinate system}
 {Laboratory coordinate system}
 {E-H-Lab}
 {The definitions of
  the (dark--green) horizontal (a)
  and laboratory (b) coordinate systems.
  $\thetaLab$ and $\phiLab$ indicate
  the latitude and the longitude of
  the location of the laboratory of interest,
  respectively.
  $\omega \tPM$ indicates
  the rotation angle of the (yellow) Prime Meridian
  from (UTC) 12 o'clock midnight (the beginning) of
  each single Solar day.
  As a reference,
  our (light--green)
  Earth coordinate system
  is also sketched here.
  See the text for the detailed descriptions.%
  }

 In Figs.~\ref{fig:E-H-Lab},
 we sketch
 the definitions of
 the (dark--green) horizontal (a)
 and laboratory (b) coordinate systems,
 respectively.
 Our (light--green)
 Earth coordinate system
 is also sketched
 as a reference here.

 Firstly,
 the origin of the horizontal coordinate system
 is chosen as the location of
 the laboratory of interest
 at (UTC) 12 o'clock midnight (the beginning) of
 each single Solar day
 with
 $\thetaLab$ and $\phiLab$ indicating
 the latitude and the longitude of
 the laboratory location,
 respectively.
 The primary direction (the $\xH$--axis)
 and the $\zH$--axis
 point towards north
 and the zenith,
 respectively.
 Then,
 as usual,
 the right--handed convention is used
 for defining the $\yH$--axis.
 Note that,
 as the Earth coordinate system,
 for each single (Solar) day,
 our horizontal coordinate system
 is fixed with the direction of the Prime Meridian
 at (UTC) 12 o'clock midnight
 and thus
 changes daily and discretely.

 Moreover,
 we consider also
 the instantaneous (UTC) incoming/scattering time of
 each recorded \CEnuNS\ event
 and define
 our laboratory coordinate system.
 It is basically the same as
 the horizontal coordinate system,
 but rotates with the considered laboratory
 around the Earth's north polar ($\zEq/\zE$--axis)
 by an angle of $\omega \tPM$,
 where
\beq
         \omega
 \equiv  \frac{2 \pi}{1~{\rm day}}
         \abrac{1 + \frac{1}{365}}
\~,
\label{eqn:omega}
\eeq
 is the angular velocity of the Earth.
 Note that
 our laboratory coordinate system changes
 (rotates around the Earth's north polar axis)
 event by event
 (see Table \ref{tab:CSs}).

\subsubsection{Transformations
               between the Earth and the horizontal coordinate systems}
\label{appx:tr-E-H}

 By definition,
 the transformations of
 the 3-D neutrino momentum
 (at the UTC incoming/scattering time $t$)
 between the Earth
 and the horizontal coordinate systems
 are pure time--independent rotations,
 which can be given by
\cheqnXa{A}
\beq
     {\bf p}_{\nu, {\rm H}}
  =  \MaEH (\thetaLab, \phiLab) \~ {\bf p}_{\nu, {\rm E}}
\~,
\label{eqn:bfpnuE->bfpnuH}
\eeq
 and,
 conversely,
\cheqnXb{A}
\beq
     {\bf v}_{\rm N_R, E}
  =  \MaHE (\thetaLab, \phiLab) \~ {\bf v}_{\rm N_R, H}
\~,
\label{eqn:bfvNRH->bfvNRE}
\eeq
\cheqnX{A}
 where
 the transformation matrices
 $\MaEH (\thetaLab, \phiLab)$ and $\MaHE (\thetaLab, \phiLab)$
 depending only on the latitude and the longitude of
 the location of the considered laboratory
 $(\thetaLab, \phiLab)$
 are given in Eqs.~(\ref{eqn:Ma_E_H}) and (\ref{eqn:Ma_H_E}).

\subsubsection{Transformations
               between the horizontal and the laboratory coordinate systems}
\label{appx:tr-H-Lab}

 Similar to Eqs.~(\ref{eqn:bfpnuE->bfpnuH})
 and (\ref{eqn:bfvNRH->bfvNRE}),
 the transformations (pure rotations) of
 the 3-D neutrino momentum
 at the UTC incoming/scattering time $t$
 between the horizontal
 and the laboratory coordinate systems
 can be given by
\cheqnXa{A}
\beq
     {\bf p}_{\nu, {\rm Lab}}
  =  \MaHLab(t, \phiLab, \thetaLab) \~ {\bf p}_{\nu, {\rm H}}
\~,
\label{eqn:bfpnuH->bfpnuLab}
\eeq
 and,
 conversely,
\cheqnXb{A}
\beq
     {\bf v}_{\rm N_R, H}
  =  \MaLabH (t, \phiLab, \thetaLab) \~ {\bf v}_{\rm N_R, Lab}
\~,
\label{eqn:bfvNRLab->bfvNRH}
\eeq
\cheqnX{A}
 where
 the transformation matrices
 $\MaHLab (t, \thetaLab, \phiLab)$ and $\MaLabH (t, \thetaLab, \phiLab)$
 depending not only on the latitude and the longitude of
 the laboratory location
 $(\thetaLab, \phiLab)$
 but also on the incoming/scattering time $t$ ($\tPM$)
 are given in Eqs.~(\ref{eqn:Ma_H_Lab}) and (\ref{eqn:Ma_Lab_H}).

\subsubsection{Transformation matrices}
\label{appx:Ma-XYZ_H-Lab}

 Firstly,
 the transformation matrices
 between the Earth
 and the horizontal coordinate systems
 can be given by
 \cite{DMDDD-N}
\cheqnXa{A}
\beqn
     \MaEH (\thetaLab, \phiLab)
 \=  \left[\begin{array}{c c c}
            -\sin(\thetaLab) \cos(\phiLab) ~&~
            -\sin(\thetaLab) \sin(\phiLab) ~&~
             \cos(\thetaLab)                \\
                             \sin(\phiLab) ~&~
            -                \cos(\phiLab) ~&~
              0                             \\
             \cos(\thetaLab) \cos(\phiLab) ~&~
             \cos(\thetaLab) \sin(\phiLab) ~&~
             \sin(\thetaLab)                \\
           \end{array}\right]
\~.
     \non\\
\label{eqn:Ma_E_H}
\eeqn
 Conversely,
 we have
\cheqnXb{A}
\beqn
     \MaHE (\thetaLab, \phiLab)
 \=  \left[\begin{array}{c c c}
            -\sin(\thetaLab) \cos(\phiLab) ~&~
                             \sin(\phiLab) ~&~
             \cos(\thetaLab) \cos(\phiLab)  \\
            -\sin(\thetaLab) \sin(\phiLab) ~&~
            -                \cos(\phiLab) ~&~
             \cos(\thetaLab) \sin(\phiLab)  \\
             \cos(\thetaLab)               ~&~
              0                            ~&~
             \sin(\thetaLab)                \\
           \end{array}\right]
\~.
     \non\\
\label{eqn:Ma_H_E}
\eeqn
\cheqnX{A}%
 Remind that,
 since,
 by our definitions,
 both of
 the Earth
 and the horizontal coordinate systems
 are fixed with the direction of the Prime Meridian
 at (UTC) 12 o'clock midnight of
 each single Solar day,
 the transformations
 between them
 depend only on the location
 (the latitude and the longitude)
 of the considered laboratory.

 Similarly,
 the transformation matrices
 between the Earth
 and the laboratory coordinate systems
 can be obtained directly as
\cheqnXa{A}
\beqn
 \conti
     \MaELab (t, \thetaLab, \phiLab)
     \non\\
 \=  \footnotesize
     \left[\begin{array}{c c c}
            -\sin(\thetaLab) \cos\abrac{\phiLab + \omega \tPM} ~&~
            -\sin(\thetaLab) \sin\abrac{\phiLab + \omega \tPM} ~&~
             \cos(\thetaLab)                                    \\
                             \sin\abrac{\phiLab + \omega \tPM} ~&~
            -                \cos\abrac{\phiLab + \omega \tPM} ~&~
              0                                                 \\
             \cos(\thetaLab) \cos\abrac{\phiLab + \omega \tPM} ~&~
             \cos(\thetaLab) \sin\abrac{\phiLab + \omega \tPM} ~&~
             \sin(\thetaLab)                                    \\
           \end{array}\right]
\~,
\label{eqn:Ma_E_Lab}
\eeqn
 and
\cheqnXb{A}
\beqn
 \conti
     \MaLabE (t, \thetaLab, \phiLab)
     \non\\
 \=  \footnotesize
     \left[\begin{array}{c c c}
            -\sin(\thetaLab) \cos\abrac{\phiLab + \omega \tPM} ~&~
                             \sin\abrac{\phiLab + \omega \tPM} ~&~
             \cos(\thetaLab) \cos\abrac{\phiLab + \omega \tPM}  \\
            -\sin(\thetaLab) \sin\abrac{\phiLab + \omega \tPM} ~&~
            -                \cos\abrac{\phiLab + \omega \tPM} ~&~
             \cos(\thetaLab) \sin\abrac{\phiLab + \omega \tPM}  \\
             \cos(\thetaLab)                                   ~&~
              0                                                ~&~
             \sin(\thetaLab)                                    \\
           \end{array}\right]
\~.
     \non\\
\label{eqn:Ma_Lab_E}
\eeqn
\cheqnX{A}%
 Then,
 by combining Eqs.~(\ref{eqn:Ma_H_E}) and (\ref{eqn:Ma_Lab_E})
 with Eqs.~(\ref{eqn:Ma_E_Lab}) and (\ref{eqn:Ma_E_H}),
 the transformation matrices
 between the horizontal
 and the laboratory coordinate systems
 can be expressed as
\cheqnXa{A}
\beqn
 \conti
     \MaHLab (t, \thetaLab, \phiLab)
     \non\\
 \=  \MaELab (t, \thetaLab, \phiLab) \~
     \MaHE      (\thetaLab, \phiLab)
     \non\\
 \=  \scriptsize
     \left[\renewcommand{\arraystretch}{2}
           \begin{array}{c c c}
                      \cos(\omega \tPM)  \sin^2(\thetaLab) + \cos^2(\thetaLab) ~&~
                      \sin(\omega \tPM)  \sin  (\thetaLab)                     ~&~
            \bbig{1 - \cos(\omega \tPM)} \sin  (\thetaLab)   \cos  (\thetaLab)  \\
                    - \sin(\omega \tPM)  \sin  (\thetaLab)                     ~&~
                      \cos(\omega \tPM)                                        ~&~
                      \sin(\omega \tPM)                      \cos  (\thetaLab)  \\
            \bbig{1 - \cos(\omega \tPM)} \sin  (\thetaLab)   \cos  (\thetaLab) ~&~
                    - \sin(\omega \tPM)                      \cos  (\thetaLab) ~&~
                      \cos(\omega \tPM)  \cos^2(\thetaLab) + \sin^2(\thetaLab)  \\
           \end{array}\right]
\~,
     \non\\
\label{eqn:Ma_H_Lab}
\eeqn
 and
\cheqnXb{A}
\beqn
 \conti
     \MaLabH (t, \thetaLab, \phiLab)
     \non\\
 \=  \MaEH      (\thetaLab, \phiLab) \~
     \MaLabE (t, \thetaLab, \phiLab)
     \non\\
 \=  \scriptsize
     \left[\renewcommand{\arraystretch}{2}
           \begin{array}{c c c}
                      \cos(\omega \tPM)  \sin^2(\thetaLab) + \cos^2(\thetaLab) ~&~
                    - \sin(\omega \tPM)  \sin  (\thetaLab)                     ~&~
            \bbig{1 - \cos(\omega \tPM)} \sin  (\thetaLab)   \cos  (\thetaLab)  \\
                      \sin(\omega \tPM)  \sin  (\thetaLab)                     ~&~
                      \cos(\omega \tPM)                                        ~&~
                    - \sin(\omega \tPM)                      \cos  (\thetaLab)  \\
            \bbig{1 - \cos(\omega \tPM)} \sin  (\thetaLab)   \cos  (\thetaLab) ~&~
                      \sin(\omega \tPM)                      \cos  (\thetaLab) ~&~
                      \cos(\omega \tPM)  \cos^2(\thetaLab) + \sin^2(\thetaLab)  \\
           \end{array}\right]
\~.
     \non\\
\label{eqn:Ma_Lab_H}
\eeqn
\cheqnX{A}%
 Note that,
 while
 the transformations
 between the Earth
 and the laboratory coordinate systems
 in Eqs.~(\ref{eqn:Ma_E_Lab}) to (\ref{eqn:Ma_Lab_E})
 depend on
 both of the laboratory location
 and
 the incoming/scattering time of
 each \CEnuNS\ event
 (recorded in the considered laboratory),
 those
 between the horizontal
 and the laboratory coordinate systems
 are longitude ($\phiLab$) independent and
 depend only on
 the latitude of the laboratory location $\thetaLab$
 and the incoming/scattering time $t$.

\begin{table} [t!]
\small
\begin{center}
\renewcommand{\arraystretch}{1.5}
\begin{tabular}{|| c || c | c || c ||}
\hline
\hline
 \makebox[4   cm][c]{Coordinate system} &
 \makebox[2.5 cm][c]{Movement}          &
 \makebox[2.5 cm][c]{Rotation}          &
 \makebox[6   cm][c]{Style}             \\
\hline
\hline
 Galactic   & $\times$          & $\times^\dagger$ & Fixed \\
\hline
 Ecliptic   & $\surd$           & $\times$         & Orbital $\to$ approximately linear \\
 Equatorial & $\surd$           & $\times$         & Linear + orbital $\to$ helical     \\
\hline
 Earth      & $\times^\ddagger$ & $\surd$          & Daily and discrete           \\
 Horizontal & $\times^\ddagger$ & $\surd$          & Daily and discrete           \\
 Laboratory & $\times^\ddagger$ & $\surd$          & Instantaneous and continuous \\
\hline
\hline
\end{tabular}
\end{center}
\caption{
 The summary of
 the styles of the movements and the rotations of
 all six celestial coordinate systems
 defined in our simulation package.
\\
 $^\dagger$:
 The tiny angle
 swept by the connection
 between the Solar and the Galactic centers
 during the orbital motion of the Solar system
 in the Galaxy
 is ignored in our package.
\\
 $^\ddagger$:
 Fixed on the Earth
 and combined additionally with
 the ``linear + orbital $\to$ helical'' movement of
 the Equatorial coordinate system.
}
\label{tab:CSs}
\end{table}
\subsection{Incoming--neutrino coordinate system}
\label{appx:XYZ_nu}

 For our MC simulation of
 3-D coherent elastic neutrino--nucleus scattering events,
 we have introduced the incoming--neutrino coordinate system
 defined in Sec.~\ref{subsubsec:XYZ_nu}.

\subsubsection{Transformations
               between the laboratory and the incoming--neutrino coordinate systems}
\label{appx:tr-Lab-nu}

 Finally,
 for the transformation (pure rotation) of
 the recoil direction of
 the neutrino--scattered target nucleus
 generated in the incoming--neutrino coordinate system
 to the laboratory coordinate system,
 one has
\beq
     {\bf v}_{\rm N_R, Lab}
  =  \ManuLab(\phinuLab, \thetanuLab) \~ {\bf v}_{\rm N_R, \nu_{in}}
\~,
\label{eqn:bfvNRnu->bfvNRLab}
\eeq
 where
 the transformation matrix
 $\ManuLab(\phinuLab, \thetanuLab)$
 depending on the elevation $\thetachiLab$
 and the azimuthal angle $\phichiLab$ of
 the moving direction of the incidnent neutrino
 measured in the laboratory coordinate system
 is given in Eq.~(\ref{eqn:Ma_nu_Lab}).

\subsubsection{Transformation matrices}
\label{appx:Ma-XYZ_nu-Lab}

 Similar to the transformations
 between the horizontal
 and the Earth coordinate systems,
 the transformation
 from the incoming--neutrino coordinate system
 to the laboratory coordinate system
 can be done by rotating
 first $\pi / 2 - \thetanuLab$ around the $\ynu$--axis
 and then $\pi     - \phinuLab$   around the $\znu$--axis
 (see Fig.~\ref{fig:nu-Lab})
 and
 can thus be given by
\beqn
 \conti
     \ManuLab (\thetanuLab, \phinuLab)
     \non\\
 \=  \left[\begin{array}{c c c}
            -\sin(\thetanuLab) \cos(\phinuLab) ~&~
                               \sin(\phinuLab) ~&~
             \cos(\thetanuLab) \cos(\phinuLab)  \\
            -\sin(\thetanuLab) \sin(\phinuLab) ~&~
            -                  \cos(\phinuLab) ~&~
             \cos(\thetanuLab) \sin(\phinuLab)  \\
             \cos(\thetanuLab)                 ~&~
              0                                ~&~
             \sin(\thetanuLab)                  \\
           \end{array}\right]
\~.
\label{eqn:Ma_nu_Lab}
\eeqn
%

%

%
%
%
 %
%
\section{Elastic WIMP--nucleus scattering spectrum}
\label{sec:dRdQ}

 In this section,
 we collect all needed formulae for
 estimating the differential event rate
 for elastic WIMP--nucleus scattering.

 The general expression for
 the differential event rate
 for elastic WIMP--nucleus scattering
 with both of
 the SI scalar
 and the SD axial--vector
 cross sections
 can be given by
 \cite{SUSYDM96, DMDDranap, Schumann19, Baudis20, Cooley21}:
\beq
     \Dd{R_{\chi {\rm N}}}{Q}
  =  \frac{\rho_0}{2 \mchi \mrN^2}
     \bbigg{  \sigma_{\chi {\rm N}}^{\rm SI} \FSIQ
            + \sigma_{\chi {\rm N}}^{\rm SD} \FSDQ  }
     \intvminQ     \bfrac{f_{1, \chi}(\vchiLab)}{\vchiLab} d\vchiLab
\~.
\label{eqn:dRdQ_SISD}
\eeq
 Here
 $R_{\chi {\rm N}}$ is
 the WIMP--scattering event rate,
 i.e.,
 the number of events
 per unit time and unit mass of detector material,
 $\rho_0$ is
 the local WIMP density near the Earth,
 $\mrN$ is
 the reduced mass of
 the WIMP mass $\mchi$
 and that of the target nucleus $\mN$,
 $f_{1, \chi}(\vchiLab)$ is
 the one--dimensional velocity distribution function of
 the WIMPs impinging on the detector,
 $\vmin (Q)$ is
 the minimal--required incoming velocity of incident WIMPs
 that can deposit the energy $Q$ in the detector:
\beq
         \vmin (Q)
  =      \sfrac{\mN}{2 \mrN^2} \~ \sqrt{Q}
\~.
\label{eqn:vmin}
\eeq

 For the radial component (magnitude) of
 the 3-D WIMP velocity distribution
 in the Equatorial/laboratory coordinate system,
 we,
 as usual,
 take into account
 the orbital motion of the Solar system around our Galaxy
 as well as
 that of the Earth around the Sun
 and thus adopt
 the shifted Maxwellian velocity distribution
 given by
 \cite{Lewin96},
 for $\vchiLab \le \vesc - \ve$,
\cheqnXa{A}
\beq
     f_{1, \sh} (\vchiLab)
  =  N_{\sh}
     \afrac{\vchiLab}{v_0 \ve}
     \bbrac{  e^{-(\vchiLab - \ve)^2 / v_0^2}
            - e^{-\vesc^2            / v_0^2} }
\~,
\label{eqn:f1v_sh_vesc_1}
\eeq
 and,
 for $\vesc - \ve \le \vchiLab \le \vesc + \ve$,
\cheqnXb{A}
\beq
     f_{1, \sh} (\vchiLab)
  =  N_{\sh}
     \afrac{\vchiLab}{v_0 \ve}
     \bbrac{  e^{-(\vchiLab - \ve)^2 / v_0^2}
            - e^{-\vesc^2            / v_0^2} }
\~,
\label{eqn:f1v_sh_vesc_2}
\eeq
\cheqnX{A}
 with the normalization constant
\beq
     N_{\sh}
  =  \bbrac{  \sqrt{\pi} \~
              \erf{\D \afrac{\vesc}{v_0}}
            - \afrac{2 \vesc}{v_0}
              e^{-\vesc^2 / v_0^2}        }^{-1}
\~.
\label{eqn:N_sh_vesc}
\eeq
 Here
 $v_0 \simeq 220$ km/s is
 the Solar orbital speed around the Galactic center
 and
 $\ve$ is
 the time--dependent Earth's velocity in the Galactic frame
 \cite{Freese88,
       SUSYDM96}:
\beq
     \ve(t)
  =  v_0 \bbrac{1.05 + 0.07 \cos\afrac{2 \pi (t - \tp)}{1~{\rm yr}}}
\~,
\label{eqn:ve}
\eeq
 with $\tp \simeq$ June 2nd,
 as the date
 on which
 the velocity of the Earth relative to the WIMP halo
 is maximal%
\footnote{
 In this paper,
 the time dependence of the Earth's velocity in the Galactic frame,
 the second term of $\ve(t)$
 has been ignored
 and
 $\ve = 1.05 \~ v_0$
 is used.
}.
 In addition,
 the escape velocity from our Galaxy
 in the position of the Solar system
 has been set as $\vesc = 550$ km/s.
 From Eqs.~(\ref{eqn:f1v_sh_vesc_1})
 and       (\ref{eqn:f1v_sh_vesc_2}),
 one can obtain that,
  for $\vmin (Q) \le \vesc - \ve$,
\cheqnXa{A}
\beqn
 \conti
     \int_{\vmin (Q)}^{\vesc + \ve}
     \bfrac{f_{1, \sh} (\vchiLab)}{\vchiLab}
     d\vchiLab
     \non\\
 \=  N_{\sh}
     \cbrac{  \frac{\sqrt{\pi}}{2 \ve}
              \bbrac{  \erf{\D \afrac{\vmin (Q) + \ve}{v_0}}
                     - \erf{\D \afrac{\vmin (Q) - \ve}{v_0}} }
            - \afrac{2}{v_0}
              e^{-\vesc^2 / v_0^2}                              }
\~,
\label{eqn:Int_f1v_v_sh_vesc_vmax_1}
\eeqn
 and,
  for $\vesc - \ve \le \vmin (Q) \le \vesc + \ve$,
\cheqnXb{A}
\beqn
 \conti
     \int_{\vmin (Q)}^{\vesc + \ve}
     \bfrac{f_{1, \sh} (\vchiLab)}{\vchiLab}
     d\vchiLab
     \non\\
 \=  N_{\sh}
     \cbrac{  \frac{\sqrt{\pi}}{2 \ve}
              \bbrac{  \erf{\D \afrac{\vesc}          {v_0}}
                     - \erf{\D \afrac{\vmin (Q) - \ve}{v_0}} }
            - \afrac{\vesc + \ve - \vmin (Q)}{v_0 \ve}
              e^{-\vesc^2 / v_0^2}                              }
\~.
\label{eqn:Int_f1v_v_sh_vesc_vmax_2}
\eeqn
\cheqnX{A}

 The SI scalar WIMP--nucleus cross section
 in Eq.~(\ref{eqn:dRdQ_SISD})
 can be expressed as
 \cite{SUSYDM96}
\beqn
     \sigma_{\chi {\rm N}}^{\rm SI}
 \=  \afrac{4}{\pi} \mrN^2 \bBig{(A - Z) \frmn + Z \frmp}^2
     \non\\
 \eqnsimeq
     A^2 \afrac{\mrN}{\mrp}^2 \sigmapSI
\~.
\label{eqn:sigma0SI}
\eeqn
 Here
 $\mrp$ is
 the reduced mass of the WIMP mass
 and the proton mass,
 $f_{\rm (n, p)}$ are
 the effective scalar couplings of WIMPs
 on neutrons and on protons,
 respectively,
 the theoretical prediction that
 the scalar couplings
 on neutrons and on protons
 are approximately equal:
\(
         \frmn
 \simeq  \frmp
\)
 has been adopted,
 the tiny mass difference
 between a neutron and a proton
 has been neglected,
 and
\beq
     \sigmapSI
  =  \afrac{4}{\pi} \mrp^2 |\frmp|^2
\~,
\label{eqn:sigmapSI}
\eeq
 is the SI WIMP--nucleon cross section.
 On the other hand,
 the SD axial--vector WIMP--nucleus cross section
 in Eq.~(\ref{eqn:dRdQ_SISD})
 can be expressed as
 \cite{SUSYDM96}
\beqn
     \sigma_{\chi {\rm N}}^{\rm SD}
 \=  \afrac{32}{\pi} G_F^2 \~ \mrN^2
     \afrac{J + 1}{J} \bBig{\Srmn \armn + \Srmp \armp}^2
     \non\\
 \=  \frac{4}{3} \afrac{J + 1}{J} \afrac{\mrN}{\mrp}^2
     \bbrac{\Srmn \afrac{\armn}{\armp} + \Srmp}^2
     \sigmapSD
\~.
\label{eqn:sigma0SD}
\eeqn
 Here
 $J$ is
 the total spin of the target nucleus
 (see Table \ref{tab:Sp/n}),
 $a_{\rm (n, p)}$ are
 the effective axial--vector WIMP couplings
 on neutrons and on protons,
 respectively,
 and
 the SD WIMP cross section
 on neutrons and on protons
 can be given as
\beq
     \sigma_{\chi {\rm (n, p)}}^{\rm SD}
  =  \afrac{24}{\pi} G_F^2 \~ m_{\rm r, (n, p)}^2 |a_{\rm (n, p)}|^2
\~,
\label{eqn:sigmap/nSD}
\eeq
 respectively.

 In our simulations
 presented in this paper,
 the SI WIMP--nucleon cross section
 has been fixed as $\sigmapSI = 10^{-12}~{\rm pb} = 1~{\rm yb}$,
 then
 the SD WIMP--neutron/proton couplings
 have been tuned as $\armn = 1.4 \times 10^{-4}$
 and $\armp = 2 \times 10^{-4}$,
 respectively.

%

%
%
%
 %
%
\section{Estimate of the statistical uncertainty on the accumulated recoil energy}
\label{appx:sigma2}

 The statistical uncertainty on
 the recorded event number
 in the $n$--th angular/energy bin
 is
\beq
     \sigma^2(N_n)
  =  N_n
\~,
\label{eqn:sigma2_N}
\eeq
 and
 the uncertainty on
 the average recoil energy
 in the $n$--th angular/energy bin
 can be estimated by
\beqn
     \sigma^2\aBig{\Bar{Q}|_n}
 \=  \frac{1}{N_n - 1} \abigg{\Bar{Q^2}|_n - \Bar{Q}|_n^2}
     \non\\
 \=  \frac{1}{N_n (N_n - 1)}
     \bbrac{                       \sum_{i = 1}^{N_n} Q_{n, i}^2
            - \frac{1}{N_n} \abrac{\sum_{i = 1}^{N_n} Q_{n, i}}^2  }
\~.
\label{eqn:sigma2_Q}
\eeqn
 By using the standard Gaussian error propagation,
 we can obtain
 the expression for
 the statistical uncertainty on
 $\D N_n \Bar{Q}|_n = \sum_{i = 1}^{N_n} Q_{n, i}$
 as
\beqn
     \sigma^2\abrac{N_n \Bar{Q}|_n}
 \=  N_n^2        \sigma^2\aBig{\Bar{Q}|_n}
   + \Bar{Q}|_n^2 \sigma^2(N_n)
     \non\\
 \=  \frac{N_n}{N_n - 1}
     \abigg{N_n \Bar{Q^2}|_n - \Bar{Q}|_n^2}
     \non\\
 \=  \frac{N_n}{N_n - 1}
     \bbrac{                         \sum_{i = 1}^{N_n} Q_{n, i}^2
            - \frac{1}{N_n^2} \abrac{\sum_{i = 1}^{N_n} Q_{n, i}}^2  }
\~.
\label{eqn:sigma2_Q/N}
\eeqn
%

%

%
%
%
 %
%

%
%

%

%
%
%
\end{document}